\documentclass[lettersize,journal]{IEEEtran}
\usepackage{amsmath,amsfonts}
\usepackage{algorithmic}
\usepackage{algorithm}
\usepackage{academicons}
\usepackage{array}
\usepackage[caption=false,font=normalsize,labelfont=sf,textfont=sf]{subfig}
\usepackage{textcomp}
\usepackage{stfloats}
\usepackage{url}
\usepackage{verbatim}
\usepackage{graphicx}
\usepackage{cite}
\usepackage[hidelinks]{hyperref}
\begin{document}

\title{Towards Physics-informed Cyclic Adversarial Multi-PSF Lensless Imaging}



\author{Abeer Banerjee, \textit{Student Member, IEEE}, and Sanjay Singh, \textit{Senior Member, IEEE}

}

\markboth{}%
{Shell \MakeLowercase{\textit{et al.}}: A Sample Article Using IEEEtran.cls for IEEE Journals}



\maketitle

\begin{abstract}
Lensless imaging has emerged as a promising field within inverse imaging, offering compact, cost-effective solutions with the potential to revolutionize the computational camera market. By circumventing traditional optical components like lenses and mirrors, novel approaches like mask-based lensless imaging eliminate the need for conventional hardware. However, advancements in lensless image reconstruction, particularly those leveraging Generative Adversarial Networks (GANs), are hindered by the reliance on data-driven training processes, resulting in network specificity to the Point Spread Function (PSF) of the imaging system. This necessitates a complete retraining for minor PSF changes, limiting adaptability and generalizability across diverse imaging scenarios. In this paper, we introduce a novel approach to multi-PSF lensless imaging, employing a dual discriminator cyclic adversarial framework. We propose a unique generator architecture with a sparse convolutional PSF-aware auxiliary branch, coupled with a forward model integrated into the training loop to facilitate physics-informed learning to handle the substantial domain gap between lensless and lensed images. Comprehensive performance evaluation and ablation studies underscore the effectiveness of our model, offering robust and adaptable lensless image reconstruction capabilities. Our method achieves comparable performance to existing PSF-agnostic generative methods for single PSF cases and demonstrates resilience to PSF changes without the need for retraining.
\end{abstract}


\section{Introduction}\label{sec_intro}
\IEEEPARstart{I}{n} recent years, lensless imaging has emerged as a focal point of research in the inverse imaging research community that can revolutionize the computational camera market with its promise of extremely compact and inexpensive imaging solutions. The pursuit of fast and accurate reconstruction methods has been the main driving force behind the exploration of novel non-traditional approaches to capturing images among which mask-based lensless imaging is a unique solution that doesn't rely on traditional optical components. The remarkable advancements in lensless image reconstruction have been made possible by the advent of novel techniques, particularly those leveraging Generative Adversarial Networks (GANs), which have exhibited exceptional reconstruction fidelity. However, a significant challenge lies in the reliance on such data-driven training processes, which often render the resulting reconstruction network specific to the Point Spread Function (PSF) of the imaging system. Even minor alterations in the PSF necessitate retraining the network, limiting its adaptability and hindering its generalizability across different imaging scenarios. Iterative optimization-based methods, including those employing untrained neural networks for lensless imaging, offer an alternative approach by incorporating physics priors into the reconstruction process. However, the convergence of these methods takes time and they are susceptible to model mismatch \cite{zeng2021robust}, leading to the introduction of artifacts in reconstructed images. Thus, while data-driven methods hold promise in enhancing model robustness, there remains a pressing need to develop general-purpose lensless cameras capable of adapting to varying PSFs with minimal modifications to network architecture or training pipelines. 

\begin{figure}[t]
\centering
\includegraphics[width=0.48\textwidth]{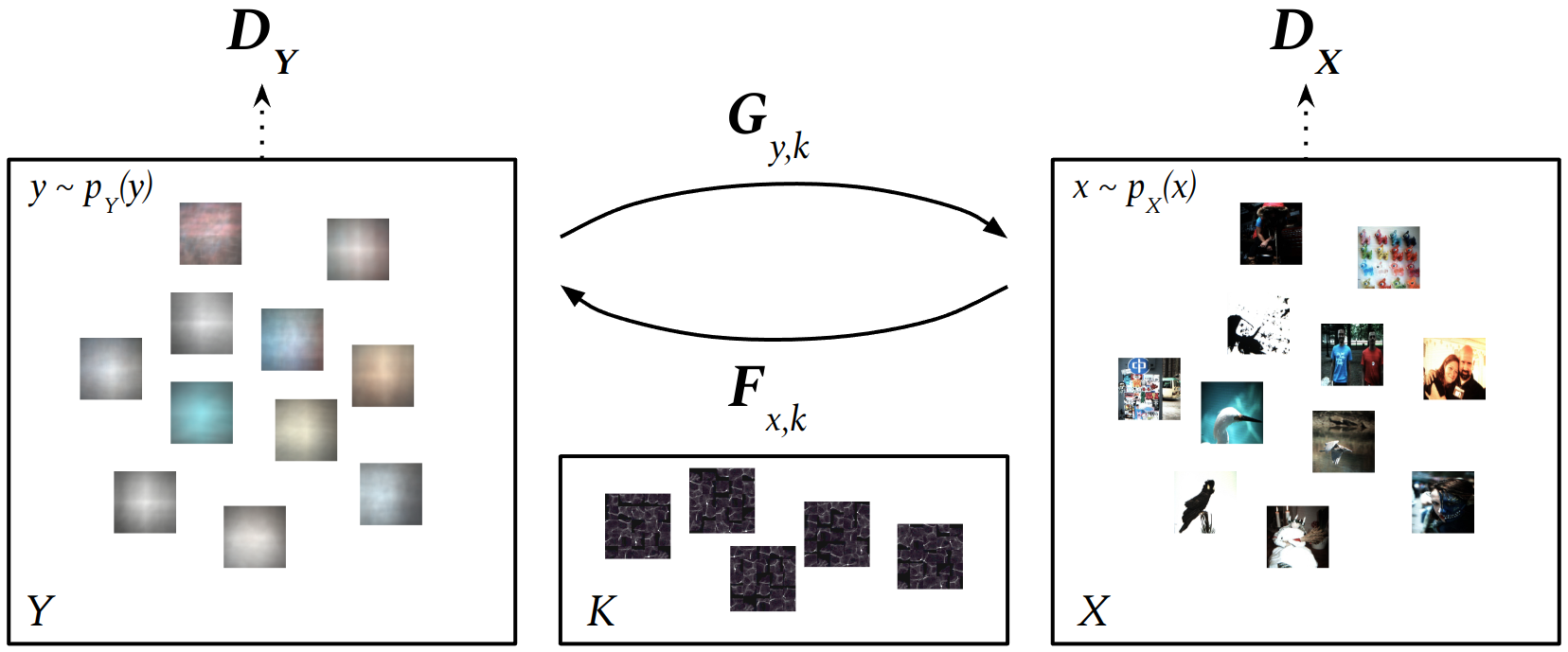}
\caption{An overview of the cyclic adversarial framework. $G_{y,k}$ refers to the generator that takes lensless image $y$ from domain $Y$ and PSF $k$ from domain $K$, as its inputs. $F_{x,k}$ refers to the forward model that takes the lensed image $x$ from domain $X$ and the PSF $k$ from domain $K$, as its inputs.}
\label{tag}
\end{figure}

Beyond the realm of research, lensless imaging holds immense potential for a diverse array of applications across various sectors. In particular, mask-based lensless imaging techniques have garnered significant attention for their versatility and applicability in fields such as biomedical imaging, environmental monitoring, and industrial inspection. The ability to capture high-resolution images using minimal hardware requirements makes mask-based lensless imaging particularly well-suited for portable diagnostic devices, remote sensing applications, and surveillance systems. Recent research has shown innovative use cases such as privacy-preserving face recognition using the concept of optical encoding \cite{wang2019privacy, henry2023privacy, wu2022lensless}

As such, bridging the gap between data-driven and physics-informed approaches in lensless imaging holds the key to unlocking its full potential across a myriad of applications. By developing robust and adaptable reconstruction methods capable of accommodating diverse PSFs, lensless imaging stands poised to redefine the boundaries of imaging technology and catalyze innovation in a wide range of fields. The PSF consists of a dark background with a sparse distribution of bright regions, known as the caustic pattern according to which the blurring effect takes place. This makes it perfectly suitable to use sparse convolutions to extract information from the PSF image. Sparse convolutions are well-suited for handling sparse data, where most of the information is concentrated in a few locations while the rest remains empty or close to zero. Since the caustic pattern in the PSF is sparsely distributed within a dark background, sparse convolutions can efficiently capture and process the relevant features without unnecessary computation on empty regions. By focusing computation only on the non-zero elements of the PSF (i.e., the caustic pattern), sparse convolutions significantly reduce the computational complexity compared to dense convolutions.

Our contributions in this paper have been summarized below:
\begin{itemize}
\item Physics-Informed Cyclic Adversarial Framework: We introduce an adversarial framework that integrates the physics of the forward model into the training loop. Our framework leverages dual discriminators corresponding to the lensless and lensed domains, ensuring robustness and fidelity in image reconstruction.

\item PSF-Aware Generator Architectures: We propose novel generator architectures designed for multi-PSF lensless imaging. Our approach encompasses both single-stage and two-stage architectures, each enabled with an auxiliary branch for handling sparse PSF input.

\item Comprehensive Performance Evaluation and Ablation Study: We conduct an extensive evaluation of our model's performance, employing both perceptual and non-perceptual evaluation metrics. We present a detailed ablation study, explaining the impact of various components and design choices on the overall reconstruction quality.
\end{itemize}

The generator $G$ is conditioned on images sampled from the lensless domain, following distribution $p_Y(y)$, and the PSF domain, following distribution $p_K(k)$. This means that $y$ is drawn from $p_{Y}(y)$ and $k$ is drawn from $p_{K}(k)$. The discriminators $D_X$ and $D_Y$ are specifically designed to operate within their respective domains of $X$ and $Y$, functioning similarly to the discriminators in the cycle GAN framework. An overview of our methodology, illustrating the translation between domains and back, is provided in Fig. \ref{tag}.

\section{Related Works}\label{sec_relw}
In the 21st century, computational imaging techniques experienced substantial growth propelled by advancements in computational capabilities, sensor technologies, and algorithmic innovations. A notable development during this period was the emergence of computational photography as a distinct field, harnessing computational methodologies to enhance and manipulate digital images in ways surpassing the limitations of traditional optical systems. Moreover, recent years have seen a transformative impact of deep learning and machine learning techniques on computational imaging. Convolutional neural networks (CNNs) and generative adversarial networks (GANs) have played pivotal roles in revolutionizing various aspects of computational imaging. These techniques have been successfully applied to a spectrum of tasks including image denoising, super-resolution, and image reconstruction, yielding remarkable outcomes and paving the way for novel avenues of exploration within the realm of computational imaging.

\subsection{Mask-based Lensless Imaging}
We can segregate the related works on lensless imaging considering two aspects of a computational camera, namely the physical layer concerning the light-capturing components in the camera, and the digital layer, concerning the algorithms being used to process the captured light and form a computational image. In the specific case of lensless imaging, which is our focus in this paper, the physical layer might consist of amplitude-modulating masks or phase-modulating masks. The binary amplitude masks allow the passage of light at certain places while blocking it in other places, giving rise to amplitude modulation. The principle although simple makes the light-capturing process lossy, with the simplest example of a binary amplitude mask being the pinhole. Approaches that have used amplitude masks for lensless imaging have achieved significant levels of reconstruction fidelity with customizations in masks such as separable amplitude pattern \cite{deweert2015lensless}, PSF-engineering via photolithography and pattern etching \cite{adams2017single, wu2020single, kuo2020chip}, Fresnel zone plates and apertures \cite{tajima2017lensless, shimano2018lensless, nakamura2016lensless, asif2016flatcam, reshetouski2020lensless}, etc. 

The binary phase grating works on the principle of phase modulation via two different heights of transparent material in front of the sensor. Although there has been significant research using phase gratings as phase modulators including works on thermal lensless imaging\cite{stork2013lensless, stork2014optical, gill2013lensless, gill2017thermal}, the imaging resolution was quite low. However, diffusers in front of sensors have proven to be an inexpensive way of achieving phase modulation and some existing lensless image reconstruction methods \cite{antipa2018diffusercam, kuo2017diffusercam} have adopted it with impressive results. In the slightly more expensive domain, fabricated custom phase masks have enabled higher reconstruction fidelity in recent methods \cite{boominathan2020phlatcam, chi2011optical}.  

\subsection{Inverse Imaging Framewroks}
Experiments at the intersection of classical methodologies and deep learning have garnered considerable attention \cite{aggarwal2018modl, zhang2017learning, dong2018denoising}. The success of these approaches has fuelled research on a wide variety of problems in inverse computational imaging giving rise to innovative frameworks that adopt learning-based strategies to address complex problems \cite{jin2017deep, ren2018learning, barbastathis2019use, sinha2017lensless, bacca2021deep}.  

Reconstruction methods for lensless imaging can be broadly categorized into iterative and non-iterative techniques, each offering distinct advantages and trade-offs. Most iterative algorithms are inherently data-agnostic in solving inverse problems. Classical MAP reconstruction algorithms, such as those based on the Alternating Direction Method of Multipliers (ADMM) and Total Variation (TV) regularization, have been extensively researched for lensless image reconstruction \cite{Rudin1992NonlinearTV, Beck2009AFI, boyd2011distributed}. However, there exists a trade-off between reconstruction quality and computational time. Non-iterative techniques, on the other hand, leverage data-driven approaches for image reconstruction, offering promising results but often requiring substantial computational resources \cite{asif2016flatcam, antipa2018diffusercam}. 

Recent paradigm-shifting experiments have demonstrated the power of representation learning in deep neural networks for solving imaging inverse problems \cite{ulyanov2018deep} showing randomly initialized neural networks can serve as effective image priors without prior training, unlike traditional neural networks that rely on large datasets for training. 
The recent advancements in physics-driven learning-based algorithms have shown impressive performance and faster convergence in a variety of inverse problems \cite{wen2023physics, poirot2019physics, deng2020interplay, karniadakis2021physics}. Untrained neural network prior-based methods have been successful at robust reconstruction by utilizing over-parameterized \cite{banerjee2022lensless, monakhova2021untrained} and under-parameterized networks \cite{banerjee2023physics} with physics-informed losses and achieved non-blind restoration. The effectiveness of deep-learning-based techniques heavily depends on the availability of large labeled datasets, which may be scarce in domains such as medical imaging and microscopy, where domain-restricted untrained reconstruction \cite{banerjee2023reconstructing} may help solve the inverse problem. In scenarios where paired datasets are available, generative adversarial networks (GANs) have emerged as powerful tools for image reconstruction, providing impressive quantitative scores and outperforming nearly all existing techniques \cite{rego2021robust, khan2019towards}. \cite{zeng2021robust} proposed a data-driven single-PSF model mismatch compensation approach in an unrolled optimization setup and improved reconstruction robustness.  

A vast majority of the existing generative methods are however PSF-agnostic leading to camera-specific networks which requires retraining once the PSF varies even slightly. In this paper, we present a novel approach for the physics-informed reconstruction of lensless images. Using the same network, we deblur lensless images corresponding to multiple point spread functions. Instead of decoupling the lensless image deblurring network from the forward model and making the reconstruction PSF-agnostic, we make our method physics-informed by incorporating multiple PSFs in the training process. This makes our method robust against not just noise but also varying PSFs. 

\section{PSF-Aware Reconstruction}
To our best knowledge, the first work on multi-PSF lensless imaging \cite{rego2021robust} solves this as a blind deblurring problem. Employing a PSF-aware method can be preferred over blind approaches that involve separate networks for PSF estimation and image reconstruction, considering that direct incorporation of the PSF into the reconstruction process allows for more accurate modeling of the image formation, leading to higher fidelity reconstructions.  

Current research lacks non-blind lensless image reconstruction approaches that can adapt to changing PSFs, and our research proposes an approach to bridge this gap. Considering that PSFs can change over time due to various factors including mechanical shifts in random phase masks, there is a need for reconfigurable networks, unlike the PSF-agnostic non-blind networks that are camera-specific.

However, measuring the PSF accurately can be challenging, and any inaccuracies in PSF measurement can degrade the reconstruction quality. Despite these challenges, the advantages of PSF-aware methods make them a promising direction for robust lensless image reconstruction in practical applications.

\section{Methodology}\label{sec_meth}
\begin{figure}[t]
\centering
\includegraphics[width=0.48\textwidth]{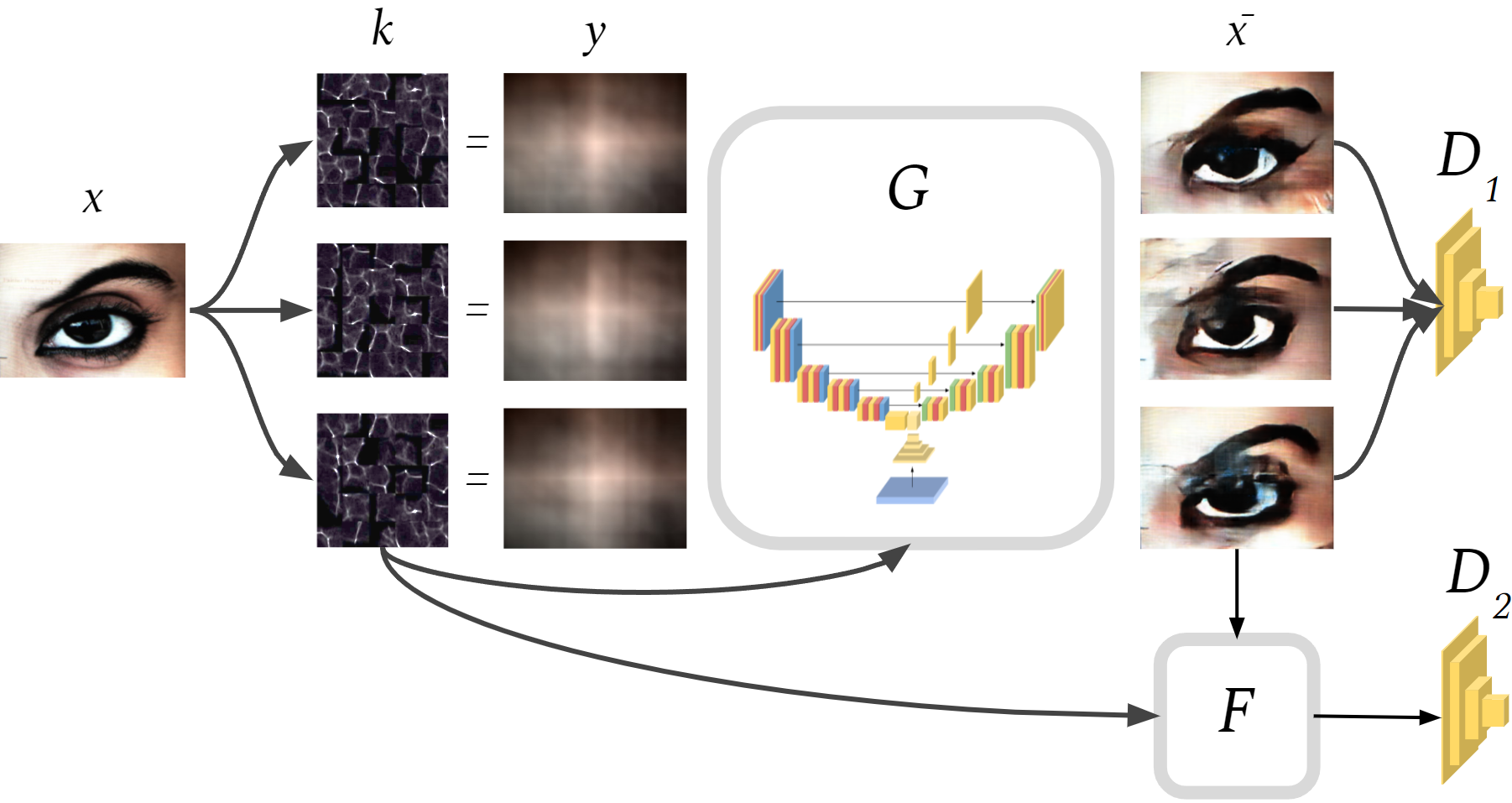}
\caption{Our dual discriminator generative adversarial reconstruction methodology for multi-PSF lensless imaging. Here $G$ is the generator, $F$ is the forward model, and $D_{1}, D_{2}$ are the two discriminators.}
\label{overv}
\end{figure}

Fig. \ref{overv} illustrates the overall workflow of our cyclic adversarial training process. Lensless images $y$ corresponding to multiple PSFs are set as input to the generator $G$ along with their corresponding PSFs $k$. We incorporate the forward model in the training loop that takes the initial prediction $\bar{x}$ and PSFs $k$ as its inputs. During the cyclic optimization process, the generator remains trainable, while the forward model does not require any training. We have extensively discussed each component of the forward model, including the architectures of the discriminators and the generator, the dataset preparation process, and the details regarding the training procedure in this section.

\subsubsection{Our Derived Dataset} We have trained our model and presented all results in this paper using a dataset derived from the benchmark dataset on lensless imaging, i.e., the DiffuserCam Lensless Mirflickr Dataset (DLMD) that comprises image pairs captured with both a lensless camera (DiffuserCam) and a conventional lensed camera \cite{monakhova2019learned}. This dataset is derived from the MIRFLICKR dataset and facilitates research in lensless imaging by being publicly available. It includes 25,000 lensless images, 25,000 lensed camera images, and the measured PSF of the lensless camera. 

We generate multiple PSFs using the seed PSF provided in the aforementioned dataset by randomly permuting sections of the seed PSF. We found that permuting 25 equal sections of the seed PSF was sufficient for generating a huge number of new PSFs, theoretically $25!$. During the training process, we use a batch of generated PSFs paired with a batch of lensed images from the DiffuserCam dataset to derive a batch of new lensless images as input by using the forward model discussed in Sec. \ref{fm}. The decision to divide the PSF into up to 25 equal sections is based on a balance between generating a substantial number of PSFs while ensuring that the pattern discontinuities introduced in the PSF generation process remain visually imperceptible. 

So, effectively, the training dataset of 25,000 images is multiplied by the number of generated PSFs each time, giving rise to a large derived training and testing dataset.

\subsubsection{The Forward Model}\label{fm} The forward model refers to the mathematical model that represents the lensless image formation process. The general equation for the forward model $\mathbb{A}$ is $y = \mathbb{A} \dot x$. In the specific inverse problem of lensless image deblurring, the lensless image formation process can be mathematically represented using a convolution operation. In this context, the lensed image $x$ is convolved with the point spread function (PSF) $k$ to obtain the lensless image $y$, thus the equation is modified to 
\begin{equation}
y = \bold{k} \ast x + \eta    
\end{equation}
In this equation, the lensed image $x$ is the input image, and the PSF $k$ represents the blurring effect caused by the absence of a lens, and $\eta$ represents the additive noise. The convolution operation effectively simulates the spreading of light from each point in the scene as it passes through the lensless system. This mathematical model assumes a linear and shift-invariant imaging system, where the blurring effect introduced by the lensless setup can be described by a fixed PSF $k$ that does not change with the location of objects in the scene. During the training process, the network prediction is in the lensed domain and is referred to as the intermediate lensed image. We computationally obtained the intermediate lensless image by convolving the intermediate lensed image with the known PSF and adding a small amount of additive Gaussian noise, thus approximating the forward process. Since the forward process was in the training loop, the approximation of the computed intermediate lensless image had to be fast and accurate, therefore we resorted to FFT convolution. With FFT convolution, the lensless image can be expressed as,
\begin{equation}
y = \mathcal{F}^{-1}(\mathcal{F}(x) \cdot \mathcal{F}(k_{new}))
\end{equation}
We add a small amount of Gaussian noise to the computed lensless image to make the reconstruction process more robust. The speed of the forward model is of paramount importance to the training pipeline since this is in the training loop and is used for calculating the cycle consistency loss.

\subsection{Network Architecture}\label{sec_arch} A CycleGAN pipeline typically consists of a pair of generators representing mappings from one domain to the other and back and a pair of discriminators for each domain. However, in the context of lensless image reconstruction, we replace the second generator that should cyclically map the prediction of the first generator back to the lensless domain with our fast and accurate approximation of the forward model. So, we only have control over the architecture of the first generator that serves as a mapping from the lensless domain to the lensed domain. A PSF-agnostic reconstruction makes the network camera-specific leading to a retraining process even for a minor shift in the PSF, therefore, we make our network PSF-aware by providing an auxiliary branch that incorporates the PSF information via sparse convolutions explained in the subsequent sections. 

\subsubsection{Dark Image Features} The Point Spread Function (PSF) is represented as a sparse tensor, where the majority of elements are zero, while only a few contain non-zero values corresponding to the illuminated regions of the caustic pattern. The characteristics of the PSF images are therefore very sparse, hence extracting meaningful features can be challenging due to the lack of prominent patterns or structures. We use two different ways to incorporate PSF information efficiently into our main network. Namely, sparse convolution and PSF subdivision approach that area elaborated in the supplementary materials section.

\begin{figure*}
\centering
\includegraphics[width=\textwidth]{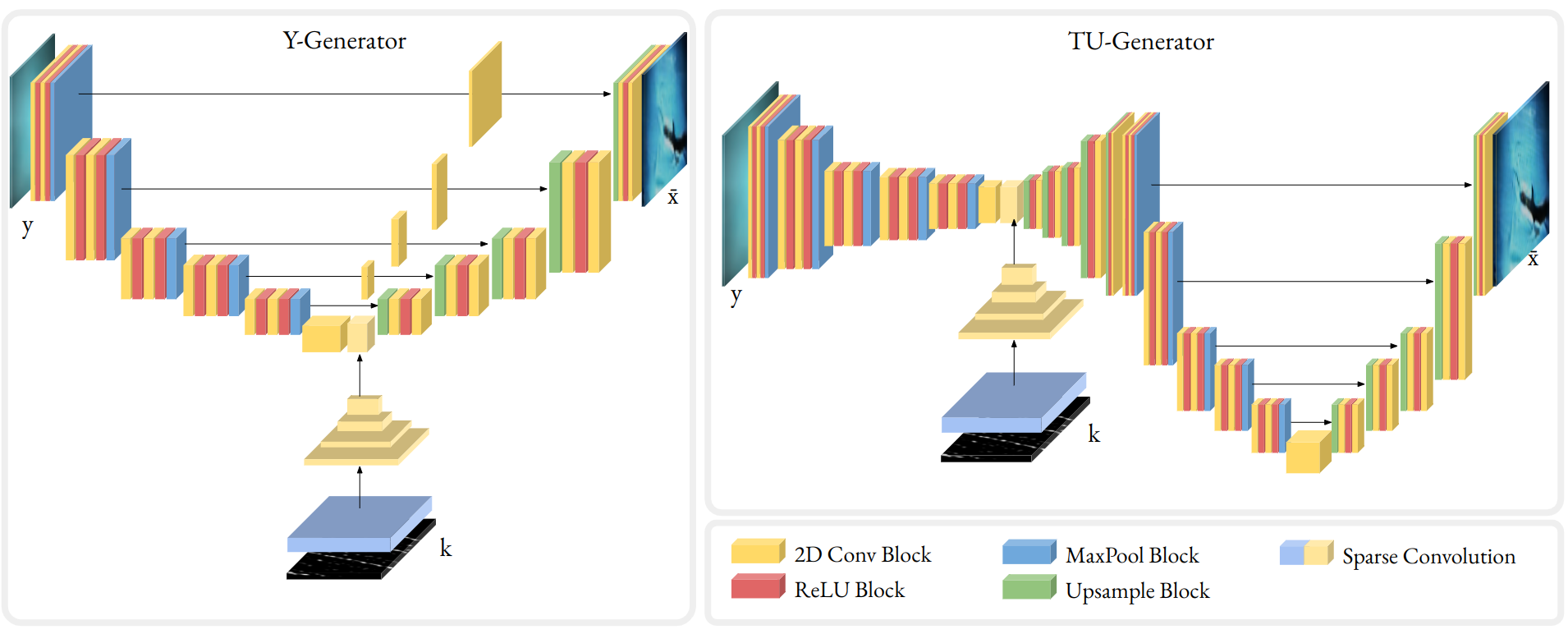}
\caption{The detailed architectures of the Y-generator and the TU-generator. In the Y-generator configuration, the CNN encoder and the decoder have skip connections, and the auxiliary sparse encoder merges at the bottleneck giving a distinct Y-shaped appearance. In the TU-generator configuration, the T-structure results due to the CNN encoder-decoder forming the upper branch and the auxiliary sparse encoder forming the stem.}
\label{arch}
\end{figure*}

\subsubsection{Generator Architectures} We have experimented with two architectures that handle the reconstruction problem in separate ways. The first variant handles the reconstruction problem in two stages. The first stage is dedicated to the physics-informed PSF-aware inversion and the second stage is for degradation restoration. The first stage comprises a CNN encoder for the lensless image and a sparse convolutional encoder for the PSF. The features computed from the lensless image and the PSF are fused and passed through a decoder consisting of upsampling and convolutional blocks. This gives the first stage its peculiar T-shaped appearance. The decoded image is then passed through a slightly modified UNet with skip connections. Owing to the structure, we call the first architectural variant TU-Generator, denoted as $G_{TU}$.

The second variant handles the reconstruction problem with a single-stage Y-shaped network. The sparse-convolutional-encoder-based PSF features and the CNN-encoder-based lensless features are computed and directly passed to a parameter-heavy decoder consisting of progressive stages of upsampling blocks followed by two convolutional blocks. The network contains skip connections from the CNN-encoder branch to the decoder branch. It can be imagined that the second variant serves as a hybrid of the first and second stages of the TU-Generator. Owing to this structure, the second variant is called the Y-Generator, denoted as $G_{Y}$. Detailed architectures of both variants of generators have been illustrated in Fig. \ref{arch}.

\subsubsection{Discriminator Architectures} Similar to the CycleGAN framework, our approach also employs two discriminators, each dedicated to one domain. We conducted experiments with various discriminator architectures and presented the ablation results in Sec. \ref{sec_abla}. Ultimately, we discovered that employing a VGG-based discriminator in the lensed domain yields favorable results, likely due to its ability to consider the entire input image as a whole, enabling the capture of global context and long-range dependencies. To implement this, we utilize a pre-trained VGG network with its fully convolutional layers, excluding the final three layers, and adapt it to output values within the range of 0 to 1, making it suitable for use as a discriminator. We denote this VGG-based discriminator as $D_{VGG}$. For the second discriminator, operating in the lensless domain, we opt for a straightforward Patch discriminator. This choice is driven by the significance of fine-grained details present in the highly multiplexed lensless images, where features are spatially localized. 
\begin{figure*}
\centering
\includegraphics[width=\textwidth]{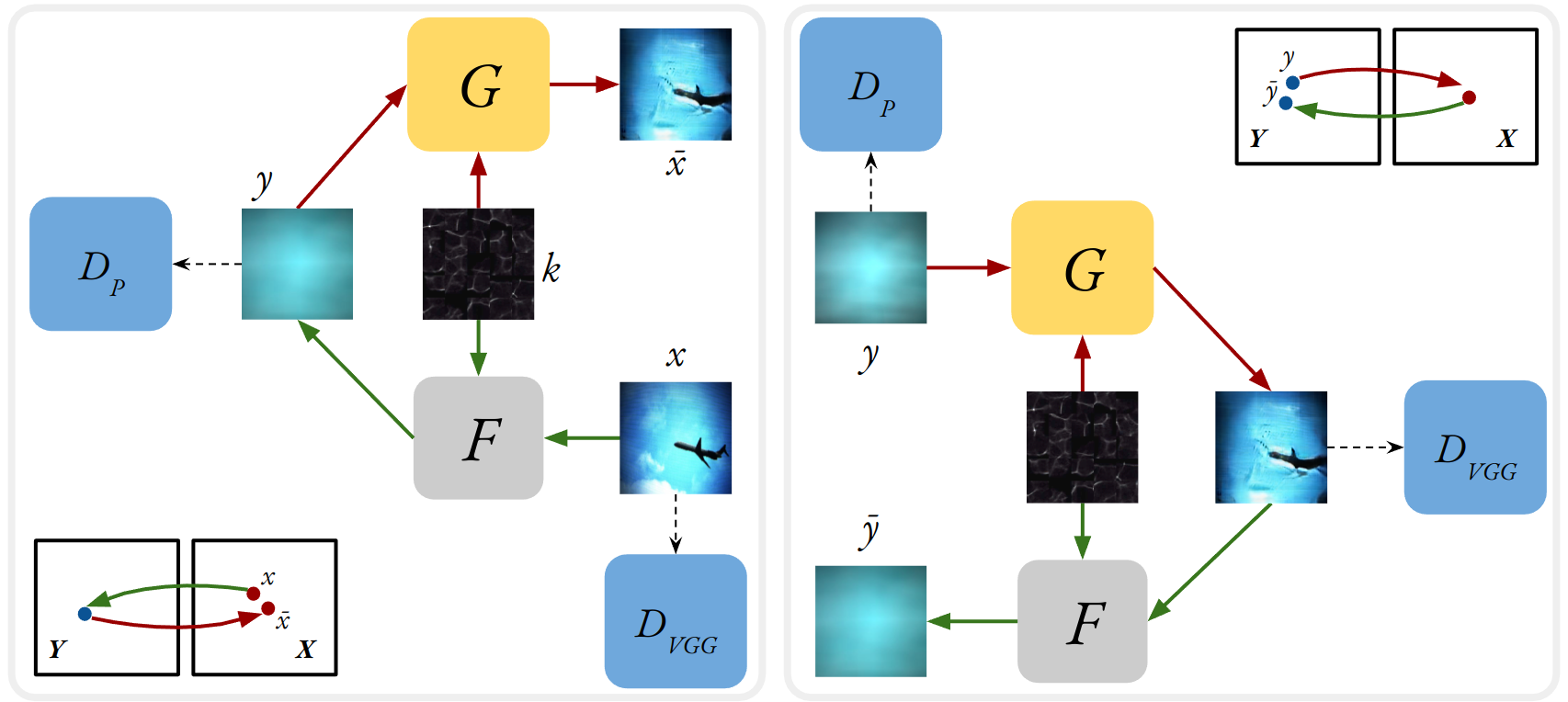}
\caption{The complete dual-discriminator physics-informed cyclic adversarial training pipeline for multi-PSF lensless imaging. Generator $G$ takes the lensless image $y$ and the PSF $k$ as its input. The forward model $F$ takes the lensed image $x$ and PSF $k$ as its input. The left half of the figure illustrates the cyclic domain translation from lensed $X$ to lensless $Y$ and back. Similarly, the right half illustrates the cyclic domain translation from lensless $Y$ to lensed $X$ and back.}
\label{trainfig}
\end{figure*}

\subsection{Loss Functions and Training}\label{sec_lft}
We frame the problem of lensless image reconstruction in a conditional GAN framework, where the hypothesis is to have a generator network as a mapping from one domain of lensless images, to another domain of lensed images, represented as $G: \mathbb{R}^{m \times n}\longrightarrow \mathbb{R}^{m \times n}$. Owing to the known PSF, we can have an approximation for the forward process that produces the lensless images given the lensed images. This helps us bring the concept of cycle consistency to the reconstruction process inspired by the CycleGAN framework. Our generator $G$ has been conditioned with the lensless input $y$ and the PSF $k$ is set as an auxiliary input. We calculate the adversarial loss using a VGG-based discriminator with the network prediction $G(y, k)$, also we support the adversarial loss with $L_1$ loss computed using the network output and the paired lensed image. We do not use a second generator that maps the prediction back to the lensless domain, instead, we use an approximation of the linear forward model to compute the lensless image, thus computing the adversarial cycle consistency loss using a Patch discriminator. The incorporation of the forward model for translating the predicted image from the lensed domain to the lensless domain involves knowledge regarding the physics of lensless image formation, thus making the GAN pipeline physics-informed. We incorporate the physics-informed consistency loss to support the adversarial loss which significantly reduces the solution space of lensed images that satisfy the inverse problem, thus helping the generator converge faster. A detailed representation of the complete dual discriminator cyclic training process has been illustrated in Fig. \ref{trainfig}.

We have experimented with two distinct variants of generator architectures to evaluate the effectiveness of our approach. The first variant of the generator architecture handles the reconstruction problem in two stages, where the initial stage performs the lensless image deblurring task and the final stage starts with the degraded inverted image and produces a perceptually improved version that resembles the lensed image. Both stages in the first variant are trained in an end-to-end fashion ensuring a robust reconstruction. The second variant is a single-stage network that is heavier in terms of parameter count compared to the first variant but concurrently handles the inversion and enhancement. The details regarding each architectural variant have been provided in Sec. \ref{sec_arch}. The overall cyclic framework corresponding to both the generators remains the same, however, the purpose of the discriminators can be changed depending on the generator configuration. An extensive ablation study has been performed and presented in Sec. \ref{sec_abla} that elaborates on the results obtained using each configuration.   

We perform the adversarial training in a Wasserstein configuration that ensures stable training dynamics since the Wasserstein distance measures the discrepancy between the true data distribution and the generated distribution more smoothly compared to vanilla counterparts, leading to stable gradients during training. This stability helps in avoiding mode collapse and gradient vanishing issues commonly encountered in traditional GAN training. 

\subsubsection{Discriminator Optimization} For each batch of size $N$ of lensless images $y_i$, lensed images $x_i$, and point spread functions $k_i$, the patch discriminator $D_P$ is used to compute $D_P(y_i)$ and the VGG disciminator $D_{VGG}$ is used to compute $D_{VGG}(x_i)$. The prediction of generator $G$ obtained using $x_i$ and $k_i$ is denoted as $\bar{x}_i = G(x_i, k_i)$, and is used to compute $D_{VGG}(\bar{x}_i)$. The predicted $\bar{x}_i$ is then passed through our forward model $F$ along with $k_i$ to generate an intermediate lensless image $\bar{y}_i = F(\bar{x}_i, k_i)$, which is then used to calculate $D_P(\bar(y)_i)$. The discriminator is then updated with the combination of these loss functions:
\begin{equation}\label{eq1}
L_{D_P}^{real} = -\frac{1}{N} \sum_{i=1}^{N} D_P(y_i) +\frac{p_d}{N} \sum_{i=1}^{N} D_P(y_i)^2    
\end{equation}

\begin{equation}\label{eq2}
L_{D_{VGG}}^{real} = -\frac{1}{N} \sum_{i=1}^{N} D_{VGG}(x_i) +\frac{p_d}{N} \sum_{i=1}^{N} D_{VGG}(x_i)^2    
\end{equation}

\begin{equation}\label{eq3}
L_{D_P}^{fake} = \frac{1}{N} \sum_{i=1}^{N} D_P(\bar{y}_i)    
\end{equation}

\begin{equation}\label{eq4}
L_{D_{VGG}}^{fake} = \frac{1}{N} \sum_{i=1}^{N} D_{VGG}(\bar{x}_i)    
\end{equation}

\begin{equation}\label{eq5}
L_{D_{P}}^{mix} = \frac{p_r}{N} \sum_{i=1}^{N} (||\nabla_b D_{P}(b)||_2 - 1)^2   
\end{equation}

\begin{equation}\label{eq6}
L_{D_{VGG}}^{mix} = \frac{p_r}{N} \sum_{i=1}^{N} (||\nabla_c D_{VGG}(c)||_2 - 1)^2   
\end{equation}

Finally, we obtain the overall discriminator loss function as:
\begin{equation}\label{eqd}
L_D = L_{D_P}^{real} + L_{D_{VGG}}^{real} + L_{D_P}^{fake} + L_{D_{VGG}}^{fake} + L_{D_P}^{mix} + L_{D_{VGG}}^{mix}
\end{equation}
We employ gradient penalty regularization by penalizing the norm of the gradient of the discriminator concerning its input. This action enforces Lipschitz continuity, leading to enhanced stability during training and improved convergence properties. To ensure the soft enforcement of the Lipschitz constraint on the discriminator, we define variables $b = \alpha \bar{y}_i + (1 - \alpha) y_i$, and $c = \alpha \bar{x}_i + (1 - \alpha) x_i$, where $\alpha \in [0,1]$. Equations for calculating the gradient penalty for the discriminators are provided in Eq. \ref{eq5} and \ref{eq6}. This controlled training process contributes to superior performance and faster convergence in image reconstruction tasks. Regulatory parameters $p_d$ and $p_r$ are introduced, with $p_d$ typically set around 0.0001, while $p_r$ is set to 1.  

\subsubsection{Generator Optimization} We use the conventional Wasserstein generator loss for optimizing the generator and support it with supervised losses. The generator loss $L_{G_W}$ is given by:
\begin{equation}\label{eqg1}
L_{G_W} = - \frac{1}{N} \sum_{i=1}^N D_{P}(\bar{y}_i) - \frac{1}{N} \sum_{i=1}^N D_{VGG}(\bar{x}_i)
\end{equation}
The supporting supervised losses $L_{G_S}$ accompanying the generator loss $L_{G_W}$ to form the final loss $L_G = \lambda_W L_{G_W} + \lambda_S L_{G_S}$, are provided below:
\begin{equation}\label{eqg2}
L_{G_S} = \frac{1}{N} \sum_{i=1}^N | x_i - \bar{x}_i | + \frac{1}{N} \sum_{i=1}^N | y_i - \bar{y}_i | 
\end{equation}
The weights $\lambda_W$ and $\lambda_S$ are adjusted according to the amount of support required by the traditional generator loss. In our case, we set $\lambda_W$ at 10 and $\lambda_S$ at 1. A detailed algorithm summarizing the optimization of the discriminators and the generator has been provided in Alg. \ref{alg1}.

\begin{algorithm*}
\caption{DDPiCF for Multi-PSF Lensless Imaging}
\label{alg1}
\begin{algorithmic}[1]
    \STATE Initialize $G$, $D_{VGG}$, and $D_P$ with random weights.
    \STATE Warm up the discriminators $D_{VGG}$ and $D_P$ for 1000 iterations by keeping the generator weights frozen.
    \FOR{k = 1,2,...} 
        \STATE Sample a batch of lensed images $\{x_1, x_2, ..., x_N\}$ from the dataset. 
        \STATE Generate a batch of PSF images $\{k_1, k_2, ..., k_N\}$ with the PSF section permutation function $\pi$.
        \STATE Generate a batch of synthetic noisy lensless images $y_i$ with the physics-informed forward model $F(x_i, k_i)$ mentioned in Sec. \ref{fm}.
        \STATE \textbf{Discriminator Update:}
        \STATE Freeze the generator $G$ and unfreeze the discriminators $D_{VGG}$ and $D_P$.
        \STATE Generate a batch of fake images $\bar{x}_i = G(y_i, k_i)$ using the generator $G$.
        \STATE Generate a batch of intermediate lensless images with the physics-informed forward model $\bar{y}_i = F(G(y_i, k_i), k_i)$.
        \STATE Compute the loss for the discriminators using the equations mentioned in Eqs. \ref{eq1}, \ref{eq2}, \ref{eq3}, \ref{eq4}, \ref{eq5}, \ref{eq6}, and finally Eq. \ref{eqd}:
            $L_D = L_{D_P}^{real} + L_{D_{VGG}}^{real} + L_{D_P}^{fake} + L_{D_{VGG}}^{fake} + L_{D_P}^{mix} + L_{D_{VGG}}^{mix}$
        \STATE Update the discriminator by minimizing the discriminator loss.

        \STATE \textbf{Generator Update:}
        \STATE Unfreeze the generator $G$ and freeze the discriminators $D_{VGG}$ and $D_P$.
        \STATE Generate a batch of synthetic noisy lensless images $y_i$ with the physics-informed forward model $F(x_i, k_i)$ mentioned in Sec. \ref{fm}. 
        \STATE Generate a batch of fake images $\bar{x}_i = G(y_i, k_i)$ using the generator $G$.
        \STATE Generate a batch of intermediate lensless images with the physics-informed forward model $\bar{y}_i = F(G(y_i, k_i), k_i)$.
        \STATE Compute the loss for Generator $G$ using Eqs. \ref{eqg1} and \ref{eqg2}, and finally compute:
            $L_G = \lambda_W L_{G_W} + \lambda_S L_{G_S}$
        \STATE Update generator by minimizing the loss.
    \ENDFOR
\end{algorithmic}
\end{algorithm*}

\section{Experiments and Results}\label{exres}
We conducted an extensive series of experiments to thoroughly evaluate the performance of our approach. Our evaluation encompasses various aspects, including the generation process of permuted PSFs, the forward model, and a comprehensive ablation study of the reconstruction pipeline. To assess the quality of our reconstructions, we employed a range of both perceptual and non-perceptual metrics. Our testing primarily focuses on images from the DiffuserCam dataset. We also performed reconstructions on a separate set of testing images in the DiffuserCam dataset to provide a fair visual comparative analysis against existing methods. This broader evaluation allows us to assess the generalizability and robustness of our method across different datasets and scenarios.

\subsection{PSF Shuffle} With the measured PSF provided in the DiffuserCam dataset, we generate multiple PSFs using a unique technique described in a prior work \cite{rego2021robust}. We generated multiple PSFs by permuting sections of the original PSF where the goal was to create a diverse set of PSFs that exhibit unique patterns. Let $k_{original}$ denote the original PSF. Dividing $k_{original}$ into $n$ equal sections, we obtain the sections $s_1, s_2,..., s_n$. A permutation of $k_{original}$ can be obtained by shuffling its sections using $\pi$ representing the permutation function that maps the indices of the sections to a new order. The new PSF can then be expressed as,
\begin{equation}
k_{new} = \pi(s_1, s_2,..., s_n)
\end{equation}
We experimentally obtained $n$ to be about 25 and by permuting these sections, we create a wide range of PSF variations which would then be convolved with the ground truth images using the forward model discussed in Sec. \ref{fm}. During the training process, the generator is conditioned with two images, the first image being the lensless image computed with the forward model with $k_{new}$ and the second being $k_{new}$ itself. The average runtime of the PSF shuffle algorithm as discussed above was measured to be 0.0002s, making it feasible to be used directly during the training process without slowing the pipeline down. The testing code will be made public on GitHub.  

\begin{figure*}[t]
\centering
\includegraphics[width=\textwidth]{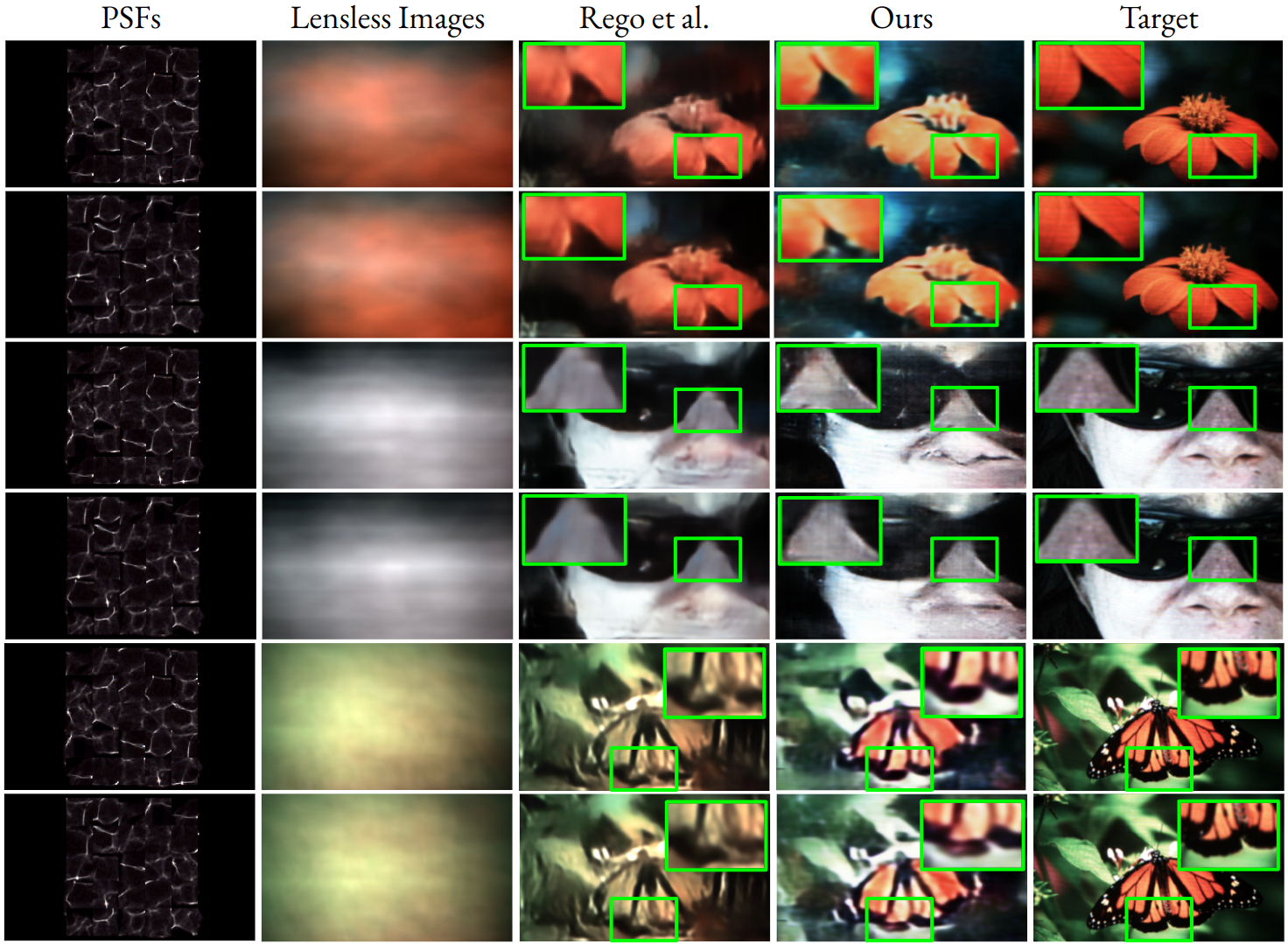}
\caption{Reconstructions corresponding to multiple-PSFs compared against Rego et al. \cite{rego2021robust}. The highlighted portions inside the green square indicate that our method can noticeably outperform the existing multi-PSF method, especially in terms of image sharpness. The Y-Net framework was used to obtain these reconstructions.}
\label{mpsfcomp}
\end{figure*}

\subsection{Quantitative Results}
We provide quantitative results using the widely accepted non-perceptual metrics of PSNR and perceptual metrics of SSIM and LPIPS. We evaluate our model on 1000 testing images that were previously isolated from the DiffuserCam dataset and compared our results against existing reconstruction frameworks corresponding to two cases: the single PSF case, and the multi-PSF case. The single PSF reconstruction results have been presented in Table \ref{tab0}.

\begin{table}[h]
\centering
\caption{Single-PSF reconstruction performance comparison of our method against the existing camera-specific methods.}
\label{tab0}
\begin{tabular}{lllll}
\hline
\hline
Method &  PSNR (dB) & SSIM & LPIPS & Time (s)\\
\hline
U-Net \cite{ronneberger2015u} & 18.69 & 0.73 & 0.265 & 0.010\\
Le-ADMM-U \cite{monakhova2019learned} & 21.16 & 0.81 & 0.203 & 0.075\\
Rego et al. \cite{rego2021robust} & 20.56 & 0.73 & - & 0.320\\
TU-Net (Ours) & 20.84 & 0.73 & 0.236 & 0.065\\
Y-Net (Ours) & 21.81 & 0.75 & 0.218 & 0.045\\ 
\hline
\hline
\end{tabular}
\end{table}
\begin{figure*}[t]
\centering
\includegraphics[width=\textwidth]{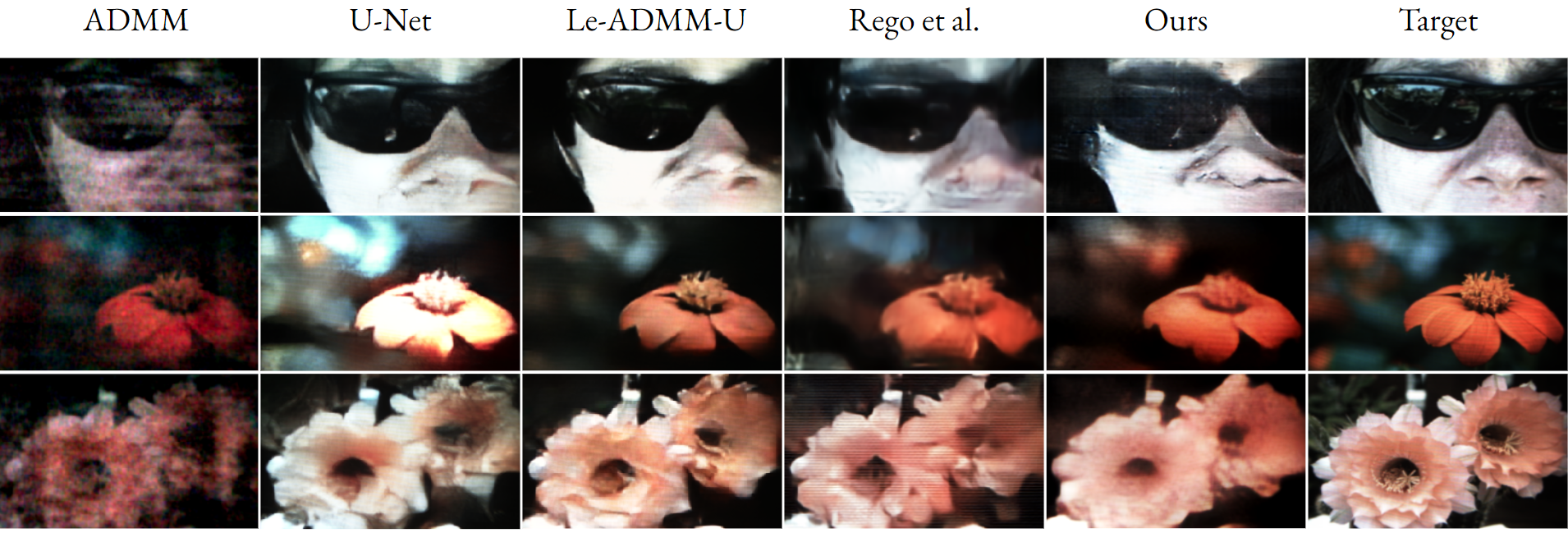}
\caption{Visual comparison of our reconstructions using the single-PSF variant of Y-Net against existing camera-specific methods. The existing methods are ADMM (iterative) \cite{boyd2011distributed}, UNet (data-driven) \cite{ronneberger2015u}, Le-ADMM-U (data-driven unrolled) \cite{monakhova2019learned}, and Rego et al. (data-driven) \cite{rego2021robust}. It can be observed that our model performs at par with the existing methods.}
\label{viscomp}
\end{figure*}

\subsection{Visual Comparison}
Quantitative metrics like PSNR and SSIM, although reliable in most cases, often are not representative of the complete picture \cite{pambrun2015limitations,ward2017image}. To offer a more informative assessment, we complement these metrics with a visual evaluation of the reconstruction results. Specifically, we compare the performance of our multi-PSF method against the existing multi-PSF method proposed by \cite{rego2021robust}. The visual comparison is provided in Fig. \ref{mpsfcomp}, where it can be observed that our reconstructions are sharper and more color-consistent across multiple PSFs. Reconstructions corresponding to multiple PSFs for other images have been provided in the supplementary section. We also compare our single-PSF variant against multiple existing camera-specific methods as illustrated in Fig. \ref{viscomp}.

\begin{figure*}
\centering
\includegraphics[width=\textwidth]{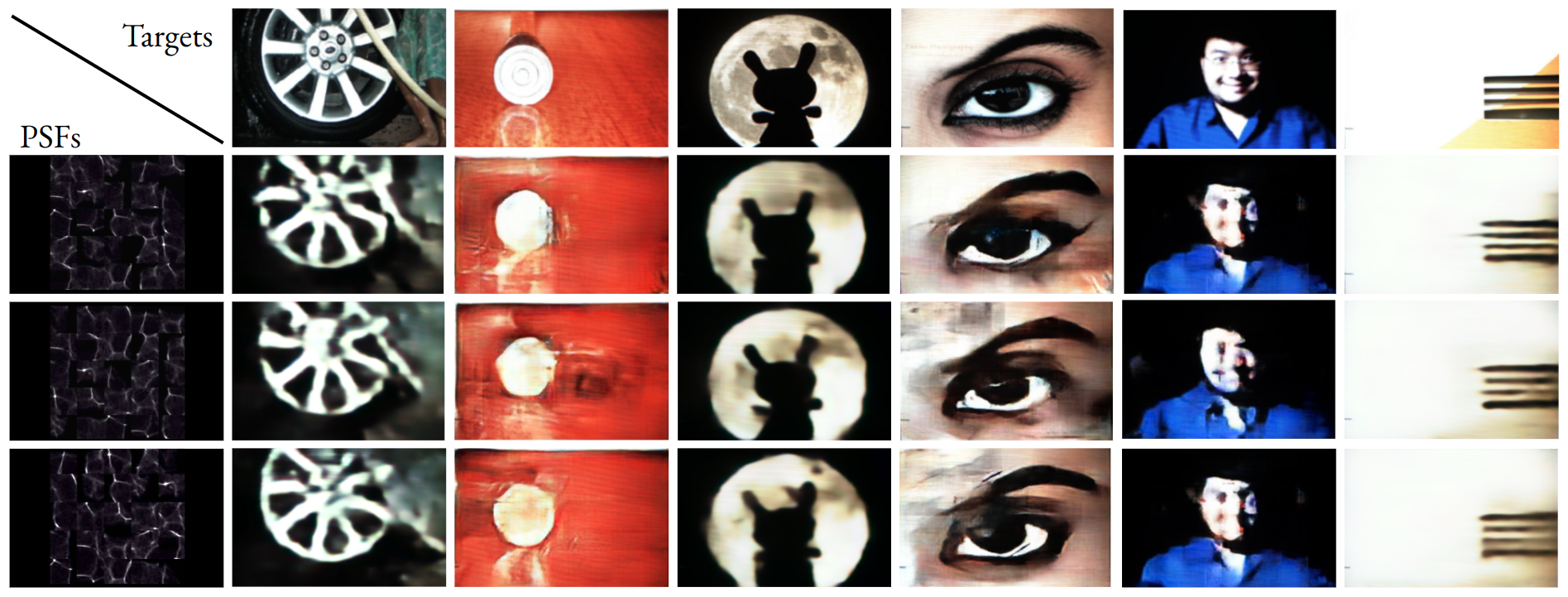}
\caption{Validation results obtained using the TU-Net architecture corresponding to multiple PSFs: It can be observed that the reconstruction quality is barely affected by changing PSFs, and the same trend is followed in the case of the Y-Net architecture.}
\label{mpsftufig}
\end{figure*}

The visual comparison of reconstructed outputs corresponding to different PSFs is presented in Fig. \ref{mpsftufig}. It can be observed that the reconstruction performance remains robust across varying PSFs, indicating the effectiveness of our approach. 


\subsection{Ablation Study}\label{sec_abla}
We conduct a thorough ablation study that explores various combinations of architectures for the generator, discriminators, and diverse loss functions. Initially, we vary the generator architecture, specifically focusing on two distinct types: the TU-Net and the Y-Net. These architectures possess contrasting structures and significantly diverge in their approaches to lensless image reconstruction. To evaluate their performance, we utilize widely accepted metrics such as PSNR, SSIM, and LPIPS, maintaining fixed discriminator architectures to enable a comparative analysis across generator variants. The outcomes are detailed in Table \ref{table2}.
\begin{table}[h]
\centering
\caption{ Multi-PSF Reconstruction performance of the generator architecture variants used in this paper.}
\label{table2}
\begin{tabular}{lllll}
\hline
\hline
Method & PSNR (dB) & SSIM & LPIPS & Time (s)\\
\hline
TU-Net-sparse & 19.19 & 0.73 & 0.269 & 0.065\\
TU-Net-unfold & 19.17 & 0.76 & 0.247 & 0.063\\
Y-Net-sparse & 20.06 & 0.75 & 0.243 & 0.048\\
Y-Net-unfold & 20.84 & 0.78 & 0.225 & 0.045\\
\hline
\hline
\end{tabular}
\end{table}
Subsequently, we perform an ablation study concerning discriminator architectures keeping the generator architecture fixed at Y-Net. Given the dedicated roles of the discriminators in the lensless and lensed domains, we explore variations encompassing patch and VGG discriminators. Our findings indicate that the VGG discriminator excels in the lensed domain owing to the effective capture of global context, while the patch discriminator proves proficient in the lensless domain by effectively discerning fine features present in the highly-multiplexed lensless images. The results of the ablation are presented in Table \ref{tab3}, and while obtaining the results, we fixed the generator architecture to be Y-Net-sparse.   
\begin{table}[h]
\centering
\caption{Multi-PSF Reconstruction performance with different combinations of discriminator pairs. The Generator was fixed to be Y-Net.}
\label{tab3}
\begin{tabular}{llll}
\hline
\hline
Discriminator 1 & Discriminator 2 & PSNR (dB) & SSIM\\
\hline
VGG-D & VGG-D & 17.59 & 0.69\\
Patch-D & VGG-D & 18.24 & 0.71\\
Patch-D & Patch-D & 18.43 & 0.71\\
VGG-D & Patch-D & 20.81 & 0.73\\
\hline
\hline
\end{tabular}
\end{table}

All computations were performed using an Intel Xeon Silver 4114 10-core CPU with 48GB RAM integrated with an NVIDIA RTX 3090 GPU.
\begin{figure}[h]
\centering
\includegraphics[width=0.48\textwidth]{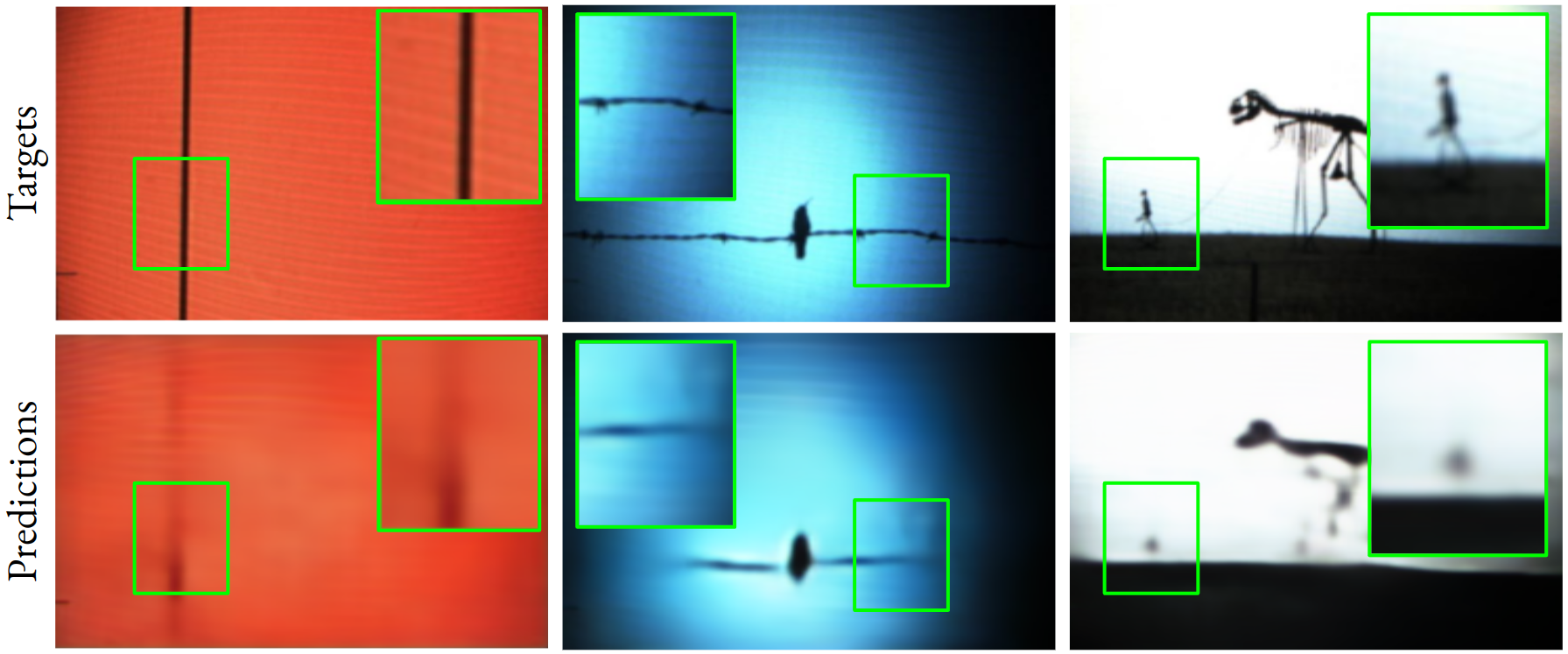}
\caption{The overall spatial and chromatic information was recovered by our method, however, the fine details could not be recovered. These reconstructions were performed with the Y-Net architecture.}
\label{fails}
\end{figure}

\section{Discussion}\label{disc}
Our methodology holds promise for application in diverse inverse imaging tasks, provided that the forward imaging process is well-defined and can be approximated in a computationally efficient way facilitating swift computation of intermediate forward computational images within the training loop. While our network's single-PSF variant demonstrates performance on par with camera-specific methods, our primary focus in this study lies in showcasing the superiority of our approach in multi-PSF scenarios. Quantitative assessments and visual comparisons underscore the efficacy of our method, highlighting its ability to surpass existing techniques in handling multiple PSFs.

\subsection{Future Work} After careful observation of the reconstruction performance, we have identified a few aspects in the pipeline where further work would be required to improve the reconstructed image. The current model although offers promising results, is limited by the resolution of the lensless input image. The physics-informed forward model receives the intermediate lensed image at a low resolution, so the computed intermediate lensless image via that model lacks fine details. The solution to this might be to use progressive GANs for improved output image resolution. Fig. \ref{fails} highlights a few instances where our model could recover the overall abstract spatial information and the chromatic information correctly, however, the fine details in the picture, including thin lines and small shapes were missed. 

\section{Conclusion}\label{conc}   
Our study presents a novel framework for multi-PSF lensless imaging, employing a dual discriminator cyclic adversarial learning strategy. Notably, our method can be distinguished from conventional cycle GAN approaches in the way that we leverage two discriminators alongside a single generator, leveraging the physics of the forward imaging process to compute cycle consistency loss, hence the name - physics-informed cyclic adversarial framework. Through comprehensive experimentation, we showcased the effectiveness of our approach in reconstructing lensless images corresponding to multiple PSFs, producing noticeably superior outcomes compared to existing techniques. Remarkably, our method showed robustness in localizing basic structural and chromatic information even for PSFs unseen during training. Evaluation using multiple perceptual and non-perceptual metrics revealed our approach to be on par with state-of-the-art camera-specific methods for single-PSF reconstruction while demonstrating a superior performance compared to the existing multi-PSF method \cite{rego2021robust}, which to our best knowledge, is the only work that addresses the problem of generalized multi-PSF lensless imaging. The potential of our approach suggests the immense scope for advancements in multi-PSF lensless image reconstruction. Future endeavors, as outlined in Section \ref{disc}, should aim to further explore and refine our framework, enabling continued innovation and progress in the field of multi-PSF lensless imaging.    

\bibliographystyle{ieeetr}
{\bibliography{bib}}

\begin{thebibliography}{10}

\bibitem{zeng2021robust}
T.~Zeng and E.~Y. Lam, ``Robust reconstruction with deep learning to handle model mismatch in lensless imaging,'' {\em IEEE Transactions on Computational Imaging}, vol.~7, pp.~1080--1092, 2021.

\bibitem{wang2019privacy}
Z.~W. Wang, V.~Vineet, F.~Pittaluga, S.~N. Sinha, O.~Cossairt, and S.~Bing~Kang, ``Privacy-preserving action recognition using coded aperture videos,'' in {\em Proceedings of the IEEE/CVF Conference on Computer Vision and Pattern Recognition Workshops}, pp.~0--0, 2019.

\bibitem{henry2023privacy}
C.~Henry, M.~S. Asif, and Z.~Li, ``Privacy preserving face recognition with lensless camera,'' in {\em ICASSP 2023-2023 IEEE International Conference on Acoustics, Speech and Signal Processing (ICASSP)}, pp.~1--5, IEEE, 2023.

\bibitem{wu2022lensless}
M.-H. Wu, Y.-T.~C. Lee, and C.-H. Tien, ``Lensless facial recognition with encrypted optics and a neural network computation,'' {\em Applied Optics}, vol.~61, no.~26, pp.~7595--7601, 2022.

\bibitem{deweert2015lensless}
M.~J. DeWeert and B.~P. Farm, ``Lensless coded-aperture imaging with separable doubly-toeplitz masks,'' {\em Optical Engineering}, vol.~54, no.~2, pp.~023102--023102, 2015.

\bibitem{adams2017single}
J.~K. Adams, V.~Boominathan, B.~W. Avants, D.~G. Vercosa, F.~Ye, R.~G. Baraniuk, J.~T. Robinson, and A.~Veeraraghavan, ``Single-frame 3d fluorescence microscopy with ultraminiature lensless flatscope,'' {\em Science advances}, vol.~3, no.~12, p.~e1701548, 2017.

\bibitem{wu2020single}
J.~Wu, H.~Zhang, W.~Zhang, G.~Jin, L.~Cao, and G.~Barbastathis, ``Single-shot lensless imaging with fresnel zone aperture and incoherent illumination,'' {\em Light: Science \& Applications}, vol.~9, no.~1, p.~53, 2020.

\bibitem{kuo2020chip}
G.~Kuo, F.~L. Liu, I.~Grossrubatscher, R.~Ng, and L.~Waller, ``On-chip fluorescence microscopy with a random microlens diffuser,'' {\em Optics express}, vol.~28, no.~6, pp.~8384--8399, 2020.

\bibitem{tajima2017lensless}
K.~Tajima, T.~Shimano, Y.~Nakamura, M.~Sao, and T.~Hoshizawa, ``Lensless light-field imaging with multi-phased fresnel zone aperture,'' in {\em 2017 IEEE International Conference on Computational Photography (ICCP)}, pp.~1--7, IEEE, 2017.

\bibitem{shimano2018lensless}
T.~Shimano, Y.~Nakamura, K.~Tajima, M.~Sao, and T.~Hoshizawa, ``Lensless light-field imaging with fresnel zone aperture: quasi-coherent coding,'' {\em Applied optics}, vol.~57, no.~11, pp.~2841--2850, 2018.

\bibitem{nakamura2016lensless}
Y.~Nakamura, T.~Shimano, K.~Tajima, M.~Sao, and T.~Hoshizawa, ``Lensless light-field imaging with fresnel zone aperture,'' in {\em ITE Technical Report 40.40 Information Sensing Technologies (IST)}, pp.~7--8, The Institute of Image Information and Television Engineers, 2016.

\bibitem{asif2016flatcam}
M.~S. Asif, A.~Ayremlou, A.~Sankaranarayanan, A.~Veeraraghavan, and R.~G. Baraniuk, ``Flatcam: Thin, lensless cameras using coded aperture and computation,'' {\em IEEE Transactions on Computational Imaging}, vol.~3, no.~3, pp.~384--397, 2016.

\bibitem{reshetouski2020lensless}
I.~Reshetouski, H.~Oyaizu, K.~Nakamura, R.~Satoh, S.~Ushiki, R.~Tadano, A.~Ito, and J.~Murayama, ``Lensless imaging with focusing sparse ura masks in long-wave infrared and its application for human detection,'' in {\em Computer Vision--ECCV 2020: 16th European Conference, Glasgow, UK, August 23--28, 2020, Proceedings, Part XIX 16}, pp.~237--253, Springer, 2020.

\bibitem{stork2013lensless}
D.~G. Stork and P.~R. Gill, ``Lensless ultra-miniature cmos computational imagers and sensors,'' {\em Proc. Sensorcomm}, pp.~186--190, 2013.

\bibitem{stork2014optical}
D.~G. Stork and P.~R. Gill, ``Optical, mathematical, and computational foundations of lensless ultra-miniature diffractive imagers and sensors,'' {\em International Journal on Advances in Systems and Measurements}, vol.~7, no.~3, p.~4, 2014.

\bibitem{gill2013lensless}
P.~R. Gill and D.~G. Stork, ``Lensless ultra-miniature imagers using odd-symmetry spiral phase gratings,'' in {\em Computational Optical Sensing and Imaging}, pp.~CW4C--3, Optica Publishing Group, 2013.

\bibitem{gill2017thermal}
P.~R. Gill, J.~Tringali, A.~Schneider, S.~Kabir, D.~G. Stork, E.~Erickson, and M.~Kellam, ``Thermal escher sensors: pixel-efficient lensless imagers based on tiled optics,'' in {\em Computational Optical Sensing and Imaging}, pp.~CTu3B--3, Optica Publishing Group, 2017.

\bibitem{antipa2018diffusercam}
N.~Antipa, G.~Kuo, R.~Heckel, B.~Mildenhall, E.~Bostan, R.~Ng, and L.~Waller, ``Diffusercam: lensless single-exposure 3d imaging,'' {\em Optica}, vol.~5, no.~1, pp.~1--9, 2018.

\bibitem{kuo2017diffusercam}
G.~Kuo, N.~Antipa, R.~Ng, and L.~Waller, ``Diffusercam: diffuser-based lensless cameras,'' in {\em Computational Optical Sensing and Imaging}, pp.~CTu3B--2, Optica Publishing Group, 2017.

\bibitem{boominathan2020phlatcam}
V.~Boominathan, J.~K. Adams, J.~T. Robinson, and A.~Veeraraghavan, ``Phlatcam: Designed phase-mask based thin lensless camera,'' {\em IEEE transactions on pattern analysis and machine intelligence}, vol.~42, no.~7, pp.~1618--1629, 2020.

\bibitem{chi2011optical}
W.~Chi and N.~George, ``Optical imaging with phase-coded aperture,'' {\em Optics express}, vol.~19, no.~5, pp.~4294--4300, 2011.

\bibitem{aggarwal2018modl}
H.~K. Aggarwal, M.~P. Mani, and M.~Jacob, ``Modl: Model-based deep learning architecture for inverse problems,'' {\em IEEE transactions on medical imaging}, vol.~38, no.~2, pp.~394--405, 2018.

\bibitem{zhang2017learning}
K.~Zhang, W.~Zuo, S.~Gu, and L.~Zhang, ``Learning deep cnn denoiser prior for image restoration,'' in {\em Proceedings of the IEEE conference on computer vision and pattern recognition}, pp.~3929--3938, 2017.

\bibitem{dong2018denoising}
W.~Dong, P.~Wang, W.~Yin, G.~Shi, F.~Wu, and X.~Lu, ``Denoising prior driven deep neural network for image restoration,'' {\em IEEE transactions on pattern analysis and machine intelligence}, vol.~41, no.~10, pp.~2305--2318, 2018.

\bibitem{jin2017deep}
K.~H. Jin, M.~T. McCann, E.~Froustey, and M.~Unser, ``Deep convolutional neural network for inverse problems in imaging,'' {\em IEEE transactions on image processing}, vol.~26, no.~9, pp.~4509--4522, 2017.

\bibitem{ren2018learning}
Z.~Ren, Z.~Xu, and E.~Y. Lam, ``Learning-based nonparametric autofocusing for digital holography,'' {\em Optica}, vol.~5, no.~4, pp.~337--344, 2018.

\bibitem{barbastathis2019use}
G.~Barbastathis, A.~Ozcan, and G.~Situ, ``On the use of deep learning for computational imaging,'' {\em Optica}, vol.~6, no.~8, pp.~921--943, 2019.

\bibitem{sinha2017lensless}
A.~Sinha, J.~Lee, S.~Li, and G.~Barbastathis, ``Lensless computational imaging through deep learning,'' {\em Optica}, vol.~4, no.~9, pp.~1117--1125, 2017.

\bibitem{bacca2021deep}
J.~Bacca, T.~Gelvez-Barrera, and H.~Arguello, ``Deep coded aperture design: An end-to-end approach for computational imaging tasks,'' {\em IEEE Transactions on Computational Imaging}, vol.~7, pp.~1148--1160, 2021.

\bibitem{Rudin1992NonlinearTV}
L.~I. Rudin, S.~Osher, and E.~Fatemi, ``Nonlinear total variation based noise removal algorithms,'' {\em Physica D: Nonlinear Phenomena}, vol.~60, pp.~259--268, 1992.

\bibitem{Beck2009AFI}
A.~Beck and M.~Teboulle, ``A fast iterative shrinkage-thresholding algorithm for linear inverse problems,'' {\em SIAM J. Imaging Sci.}, vol.~2, pp.~183--202, 2009.

\bibitem{boyd2011distributed}
S.~Boyd, N.~Parikh, E.~Chu, B.~Peleato, J.~Eckstein, {\em et~al.}, ``Distributed optimization and statistical learning via the alternating direction method of multipliers,'' {\em Foundations and Trends{\textregistered} in Machine learning}, vol.~3, no.~1, pp.~1--122, 2011.

\bibitem{ulyanov2018deep}
D.~Ulyanov, A.~Vedaldi, and V.~Lempitsky, ``Deep image prior,'' in {\em Proceedings of the IEEE conference on computer vision and pattern recognition}, pp.~9446--9454, 2018.

\bibitem{wen2023physics}
B.~Wen, S.~Ravishankar, Z.~Zhao, R.~Giryes, and J.~C. Ye, ``Physics-driven machine learning for computational imaging [from the guest editor],'' {\em IEEE Signal Processing Magazine}, vol.~40, no.~1, pp.~28--30, 2023.

\bibitem{poirot2019physics}
M.~G. Poirot, R.~H. Bergmans, B.~R. Thomson, F.~C. Jolink, S.~J. Moum, R.~G. Gonzalez, M.~H. Lev, C.~O. Tan, and R.~Gupta, ``Physics-informed deep learning for dual-energy computed tomography image processing,'' {\em Scientific reports}, vol.~9, no.~1, p.~17709, 2019.

\bibitem{deng2020interplay}
M.~Deng, S.~Li, Z.~Zhang, I.~Kang, N.~X. Fang, and G.~Barbastathis, ``On the interplay between physical and content priors in deep learning for computational imaging,'' {\em Optics Express}, vol.~28, no.~16, pp.~24152--24170, 2020.

\bibitem{karniadakis2021physics}
G.~E. Karniadakis, I.~G. Kevrekidis, L.~Lu, P.~Perdikaris, S.~Wang, and L.~Yang, ``Physics-informed machine learning,'' {\em Nature Reviews Physics}, vol.~3, no.~6, pp.~422--440, 2021.

\bibitem{banerjee2022lensless}
A.~Banerjee, H.~Kumar, S.~Saurav, and S.~Singh, ``Lensless image reconstruction with an untrained neural network,'' in {\em International Conference on Image and Vision Computing New Zealand}, pp.~430--441, Springer, 2022.

\bibitem{monakhova2021untrained}
K.~Monakhova, V.~Tran, G.~Kuo, and L.~Waller, ``Untrained networks for compressive lensless photography,'' {\em Optics Express}, vol.~29, no.~13, pp.~20913--20929, 2021.

\bibitem{banerjee2023physics}
A.~Banerjee, S.~Saurav, and S.~Singh, ``Physics-informed deep deblurring: Over-parameterized vs. under-parameterized,'' in {\em 2023 IEEE International Conference on Image Processing (ICIP)}, pp.~1615--1619, IEEE, 2023.

\bibitem{banerjee2023reconstructing}
A.~Banerjee, H.~Kumar, S.~Saurav, and S.~Singh, ``Reconstructing synthetic lensless images in the low-data regime,'' 2023.

\bibitem{rego2021robust}
J.~D. Rego, K.~Kulkarni, and S.~Jayasuriya, ``Robust lensless image reconstruction via psf estimation,'' in {\em Proceedings of the IEEE/CVF Winter Conference on Applications of Computer Vision}, pp.~403--412, 2021.

\bibitem{khan2019towards}
S.~S. Khan, V.~Adarsh, V.~Boominathan, J.~Tan, A.~Veeraraghavan, and K.~Mitra, ``Towards photorealistic reconstruction of highly multiplexed lensless images,'' in {\em Proceedings of the IEEE/CVF International Conference on Computer Vision}, pp.~7860--7869, 2019.

\bibitem{monakhova2019learned}
K.~Monakhova, J.~Yurtsever, G.~Kuo, N.~Antipa, K.~Yanny, and L.~Waller, ``Learned reconstructions for practical mask-based lensless imaging,'' {\em Optics express}, vol.~27, no.~20, pp.~28075--28090, 2019.

\bibitem{ronneberger2015u}
O.~Ronneberger, P.~Fischer, and T.~Brox, ``U-net: Convolutional networks for biomedical image segmentation,'' in {\em Medical image computing and computer-assisted intervention--MICCAI 2015: 18th international conference, Munich, Germany, October 5-9, 2015, proceedings, part III 18}, pp.~234--241, Springer, 2015.

\bibitem{pambrun2015limitations}
J.-F. Pambrun and R.~Noumeir, ``Limitations of the ssim quality metric in the context of diagnostic imaging,'' in {\em 2015 IEEE international conference on image processing (ICIP)}, pp.~2960--2963, IEEE, 2015.

\bibitem{ward2017image}
C.~M. Ward, J.~Harguess, B.~Crabb, and S.~Parameswaran, ``Image quality assessment for determining efficacy and limitations of super-resolution convolutional neural network (srcnn),'' in {\em Applications of Digital Image Processing XL}, vol.~10396, pp.~19--30, SPIE, 2017.

\end{thebibliography}

\begin{IEEEbiography}[{\includegraphics[width=1in,height=1.25in,clip,keepaspectratio]{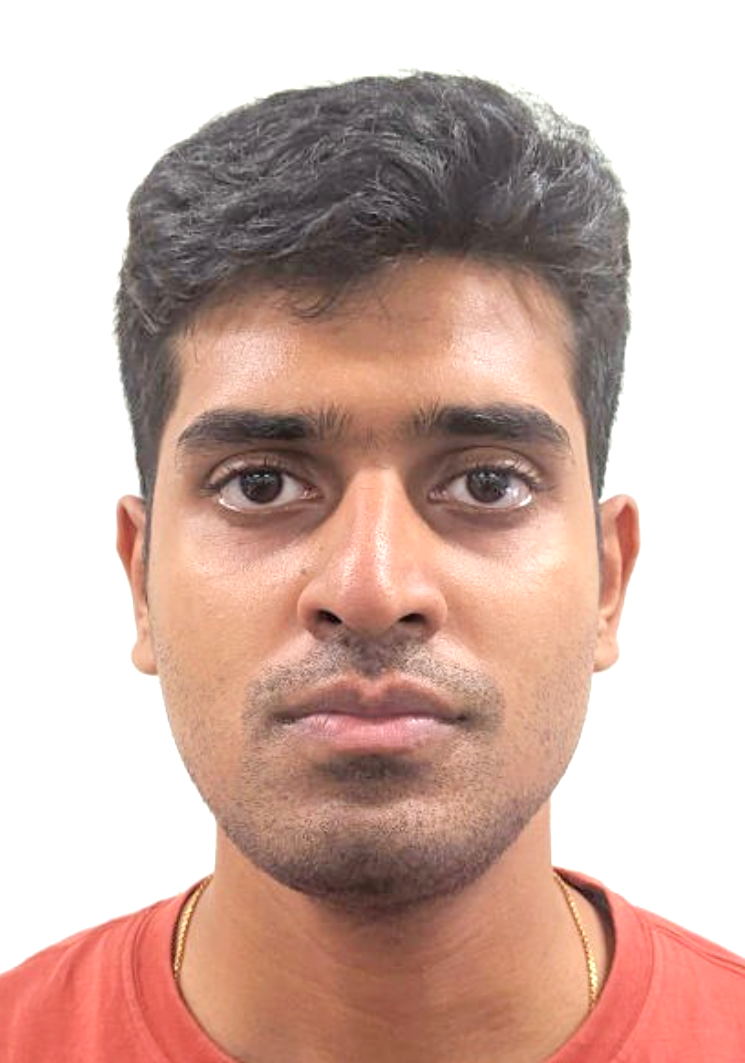}}]%
{Abeer Banerjee} obtained his B.Tech in Electronics and Communications Engineering from the University of Calcutta. Currently a Senior Research Fellow funded by CSIR-HRDG and a third-year AcSIR Integrated Dual Degree PhD student hosted by CSIR-Central Electronics Engineering Research Institute (CSIR-CEERI), his research lies at the intersection of computational imaging and machine learning, specifically exploring learning-based solutions for imaging inverse problems.  
\end{IEEEbiography}
\begin{IEEEbiography}[{\includegraphics[width=1in,height=1.25in,clip,keepaspectratio]{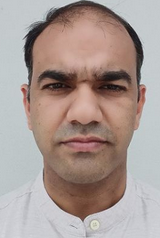}}]%
{Sanjay Singh} holds a B.Sc. in Electronics and Computer Science, an M.Sc. in Electronic Science, an M.Tech. in Microelectronics and VLSI Design, and a Ph.D. from Kurukshetra University, India. In 2009, he joined CSIR-Central Electronics Engineering Research Institute (CSIR-CEERI), Pilani as a Scientist Fellow. He has held the position of Designated Associate Professor at Hiroshima University from 2017 to 2018. Currently, he serves as a Principal Scientist and Group Head of the Advanced Information Systems Group (AITG) at CSIR-CEERI, Pilani, and as an Associate Professor at the Academy of Scientific and Innovative Research (AcSIR). His research interests span Computer Vision, Machine Learning, Artificial Intelligence, VLSI Architectures, and FPGA Prototyping.    
\end{IEEEbiography}
\vfill
\clearpage

\section{Supplementary Material}

\textbf{Sparse Convolutions:} Each non-zero element within the PSF tensor denotes a specific position or coordinate within the PSF image. To accommodate this sparse nature of the PSF, we devised a \texttt{sparsify} function tailored to transform the PSF image tensor into a sparse representation. Within this function, we identify the coordinates and their corresponding values for the illuminated pattern in the tensor. Subsequently, we pad these coordinates and values to match the size of the original input tensor. To facilitate the convolutional operations on sparse inputs, we constructed five customized convolutional layers. During the forward pass, we reshape the input sparse coordinates and values, converting them into a dense tensor where non-specified positions are filled with zeros. This dense tensor then undergoes convolutional operations with our custom layers, ultimately yielding the resulting output. This optimization significantly reduces the computational complexity of the convolution operation, making it feasible to process large PSF images efficiently. This sparse convolution block is set as an auxiliary branch to the main generator network and the sparse features are combined at the bottleneck of the generator.
\begin{figure}[h]
\centering
\includegraphics[width=0.48\textwidth]{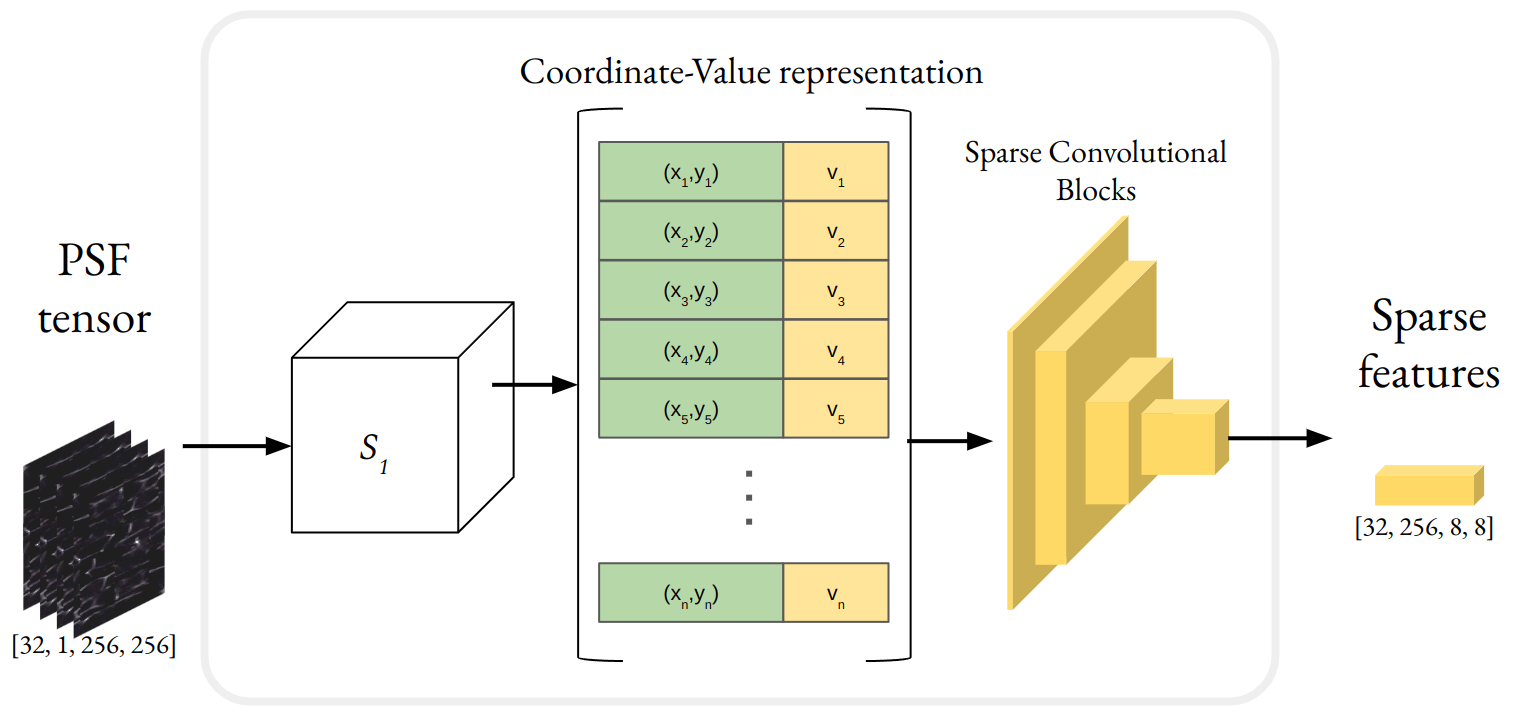}
\caption{PSF features obtained using sparse convolution. The function $S_1$ is the \texttt{sparsify} function and the function $S_2$ is the \texttt{scatter\_add} function. The PSF image tensor is set as the input to the sparsification block and we obtain the sparse features at the end.}
\label{sparse}
\end{figure}
\begin{figure}[h]
\centering
\includegraphics[width=0.48\textwidth]{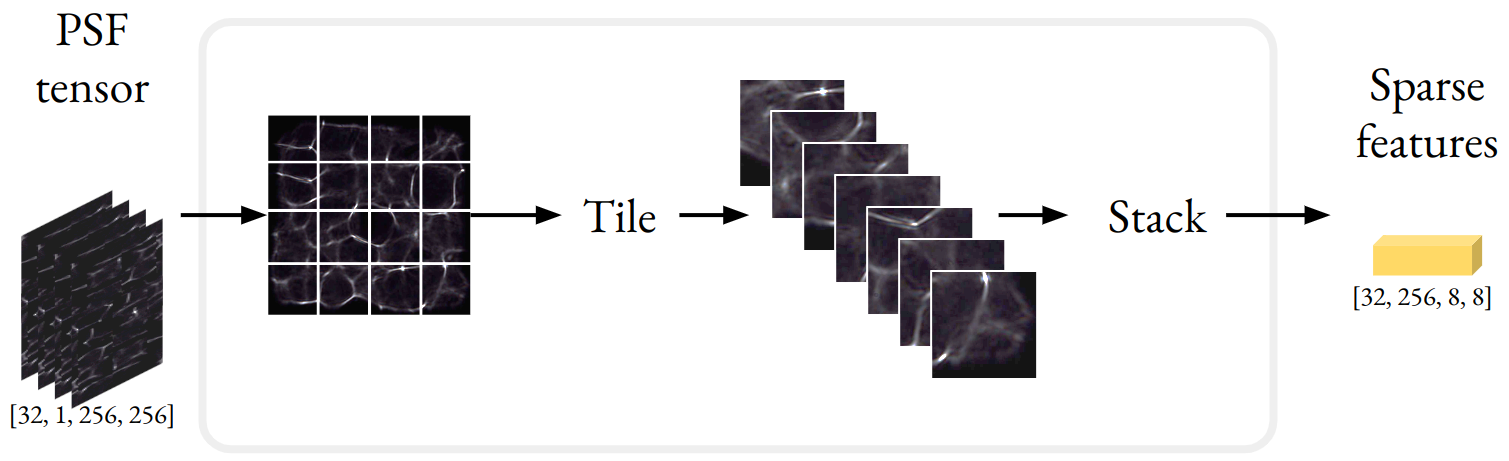}
\caption{PSF features obtained via PSF subdivision. The PSF was simply unfolded into tiles of lower spatial resolution which were then stacked to form the auxiliary PSF features.}
\label{sparse2}
\end{figure}

\textbf{PSF subdivision:} To preserve the fine features in the PSF, we subdivide the PSF into several tiles instead of directly applying any convolutional or pooling layer. We stack up the resulting tiles of the PSF contiguously and pass them through a convolutional layer. The resulting features are then passed to the main network as auxiliary information at the T-junction or the Y-junction depending on the generator architecture being used.     

\section{2D Convolution}
\begin{table}[h]
\centering
\caption{2D Convolution Runtime Analysis: The referred methods for 2D convolution were tested for a batch size of 32.}
\label{tab1}
\begin{tabular}{lll}
\hline
\hline
Method & Time (s) \\
\hline
\texttt{torch.nn.functional.conv2D} & 0.1135\\
\texttt{numpy.fft.fft2} & 0.0765\\
\texttt{scipy.signal.fftconvolve} & 0.0350\\
\hline
\hline
\end{tabular}
\end{table}
We tested various approaches for performing a fast computation of the intermediate lensless images using the approximate forward model. The FFT algorithm has a complexity of $\mathcal{O}(n \log n)$, where $n$ is the size of the input image. This is significantly faster than the $\mathcal{O}(n^2)$ complexity of direct convolution methods, especially for large inputs. Therefore, it is natural to use the forward model that computes the lensless image $y$ in the Fourier domain using the formula: $y = \mathcal{F}^{-1}(\mathcal{F}(x) \cdot \mathcal{F}(k_{new}))$. The results of the runtime analysis of 2D-convolution at different resolutions of the PSF and the lensed image averaged over 1000 runs have been presented in Table \ref{tab1}.

\begin{figure*}
\centering
\includegraphics[width=0.9\textwidth]{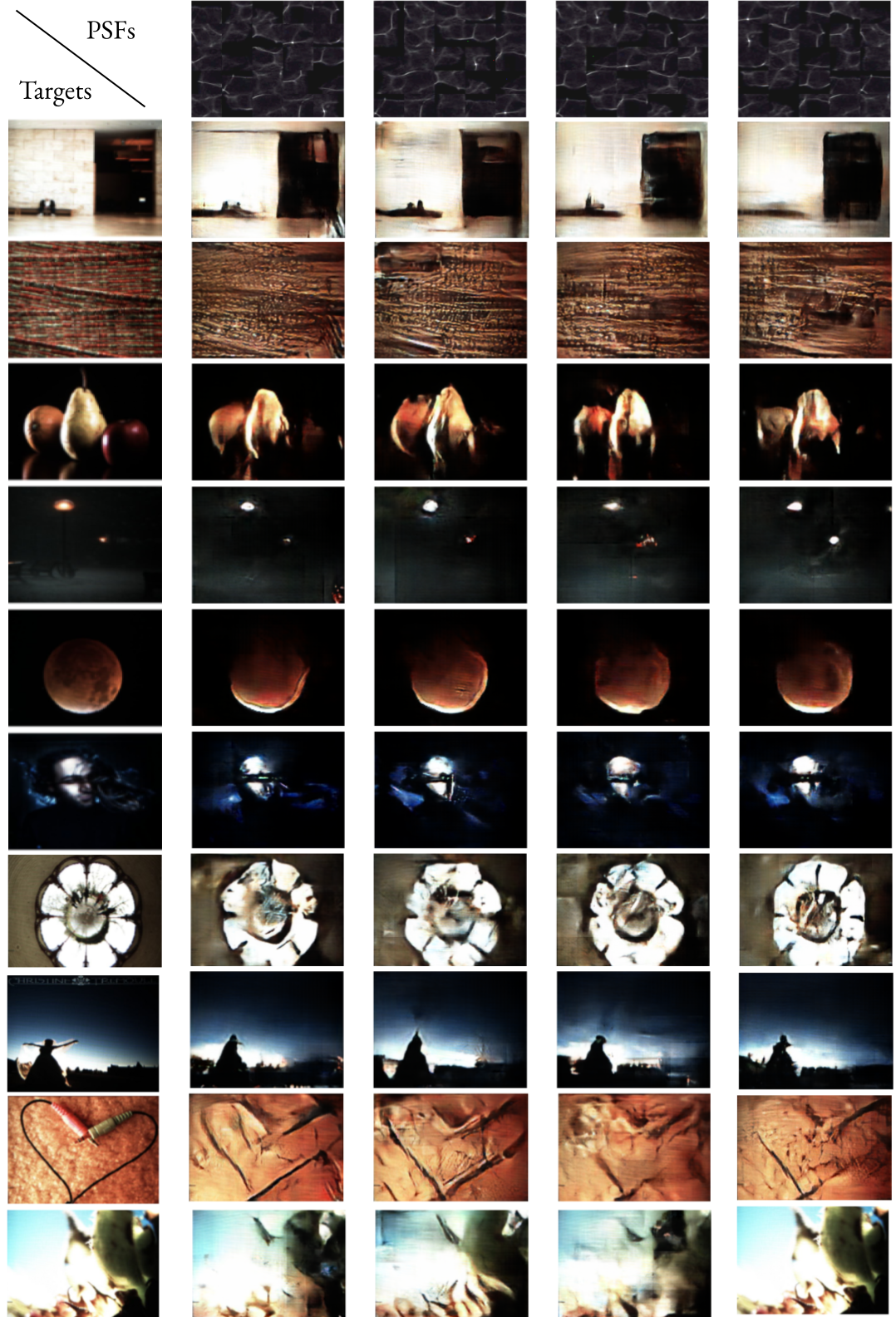}
\caption{More multi-PSF reconstruction results obtained using the Y-Net architecture.}
\label{multifig}
\end{figure*}

Figure \ref{multifig} presents additional outcomes related to multi-PSF lensless imaging. The validation results showcased in this figure demonstrate sharp reconstructions with consistent structural and chromatic performance across a range of different PSFs. These findings underscore the robustness and reliability of our imaging approach under varying conditions.\\

\begin{figure*}
\centering
\includegraphics[width=0.75\textwidth]{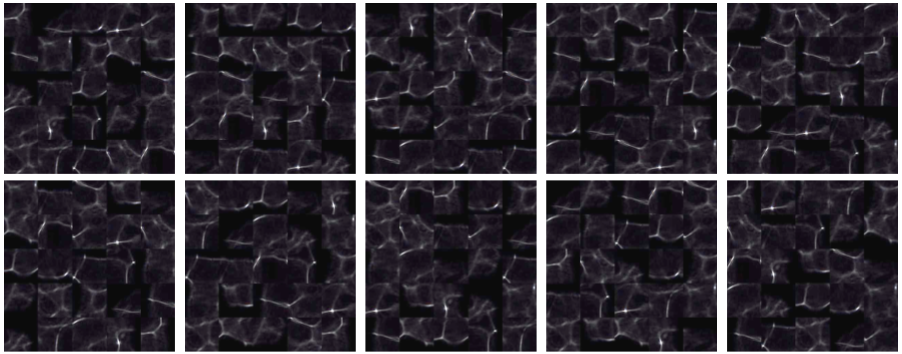}
\caption{Sample PSFs generated via PSF-shuffling technique. The generated  PSFs have been cropped and brightened for a better view.}
\label{psfs}
\end{figure*}

Figure \ref{psfs} showcases multiple instances of PSFs generated using the PSF-shuffle technique. The displayed PSFs show sufficiently diverse patterns hence proving this simple shuffling technique to be quite effective.\\

\begin{figure*}
\centering
\includegraphics[width=0.9\textwidth]{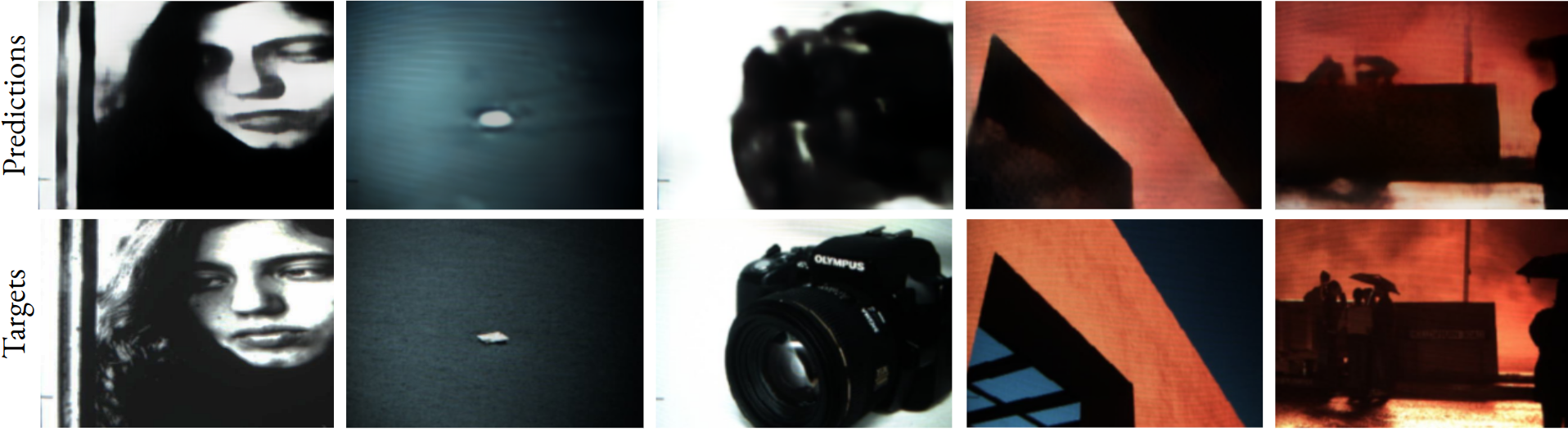}
\caption{Results generated using the Y-Net architecture trained using only MSE Loss and the physics-informed consistency loss. Notably, no discriminators were used, hence the training was non-adversarial.}
\label{cnns}
\end{figure*}

Figure \ref{cnns} presented in this figure are reconstructions produced using the Y-Net architecture, which was trained using Mean Squared Error (MSE) loss and the physics-informed consistency loss. Notably, no adversarial loss was utilized during training, so there was no need for discriminators. While the reconstructed outputs are structurally accurate, they exhibit a lack of vibrant colors and may appear somewhat unnatural. This observation sheds light on the potential impact and limitations of adversarial training in enhancing image realism.\\

\begin{figure*}
\centering
\includegraphics[width=0.9\textwidth]{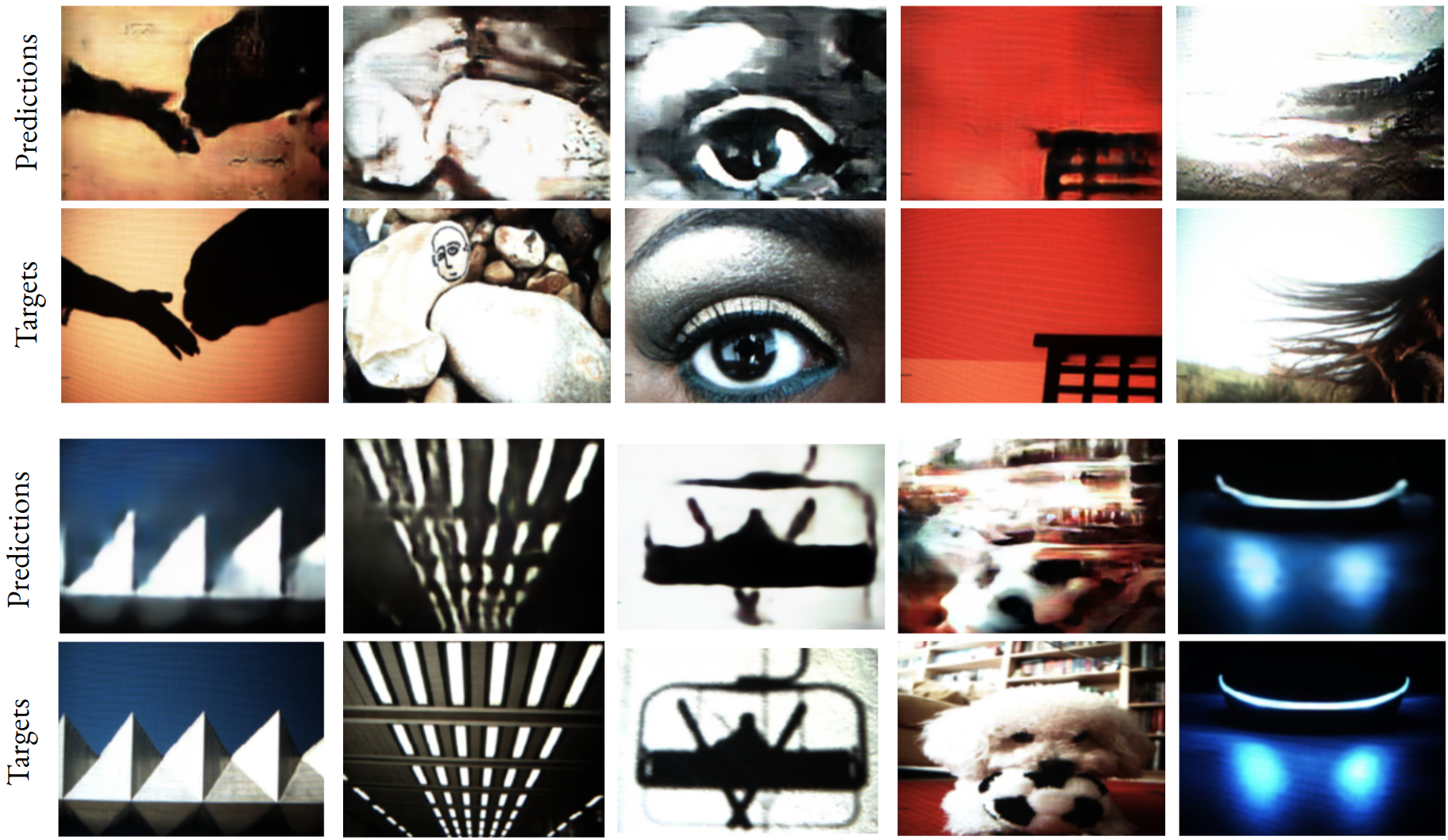}
\caption{Reconstruction results obtained using the TU-Net architecture. It can be observed that even though the outputs consist of rich colors, there are some structural artifacts still present which might be due to the insufficient resolution of PSF in the loop.}
\label{tunet}
\end{figure*}

Figure \ref{tunet} illustrates the reconstruction performance of the TU-Net architecture trained using a physics-informed cyclic adversarial pipeline. The resulting reconstructions have rich colors and a natural appearance; however, some artifacts are present, particularly in regions with fine details that are not appropriately reconstructed. Despite these artifacts, the overall visual quality and realism of the reconstructions highlight the efficacy of the proposed adversarial training strategy in improving image fidelity and appearance.\\
\\
\\

\end{document}